\documentclass[12pt,preprint]{aastex}

\usepackage{epsfig}
\usepackage{subfigure}
\usepackage{multirow}
\usepackage{amsmath,amssymb}

\newcommand{\Dnu}{\Delta \nu}
\newcommand{\dnua}{\delta \nu_\mathrm{02}}
\newcommand{\dnub}{\delta \nu_\mathrm{01}}
\newcommand{\dnuc}{\delta \nu_\mathrm{03}}
\newcommand{\muhz}{$\mu$Hz}
\newcommand{\dP}{\Delta P_\mathrm{obs}}
\newcommand{\teff}{T_\mathrm{eff}}

\newcommand{\feh}{\left[ \text{Fe/H} \right]}

\shorttitle{Asteroseismology of the open clusters NGC 6791, NGC 6811, and NGC 6819}
\shortauthors{Corsaro et al.}

\begin{document}

\title{Asteroseismology of the open clusters NGC 6791, NGC 6811, and NGC 6819 from nineteen months of \textit{Kepler} photometry}

\author{Enrico~Corsaro,\altaffilmark{1,2}
Dennis~Stello,\altaffilmark{3}
Daniel~Huber,\altaffilmark{3,4} 
Timothy~R.~Bedding,\altaffilmark{3}
Alfio~Bonanno,\altaffilmark{2}
Karsten~Brogaard,\altaffilmark{5}
Thomas~Kallinger,\altaffilmark{6}
Othman~Benomar,\altaffilmark{3}
Timothy~R.~White,\altaffilmark{3}
Benoit~Mosser,\altaffilmark{7}
Sarbani~Basu,\altaffilmark{8}
William~J.~Chaplin,\altaffilmark{9,10}
J{\o}rgen~Christensen-Dalsgaard,\altaffilmark{11,10}
Yvonne~P.~Elsworth,\altaffilmark{9}
Rafael~A.~Garc\'{\i}a,\altaffilmark{12,10}
Saskia~Hekker,\altaffilmark{13}
Hans~Kjeldsen,\altaffilmark{11}
Savita~Mathur,\altaffilmark{14,9}
S{\o}ren~Meibom,\altaffilmark{15}
Jennifer~R.~Hall,\altaffilmark{16}
Khadeejah~A.~Ibrahim,\altaffilmark{16}
and Todd~C.~Klaus\altaffilmark{16}
}
\altaffiltext{1}{Department of Physics and Astronomy, Astrophysics Section, University of Catania, Via S. Sofia 78, I-95123 Catania, Italy}
\altaffiltext{2}{I.N.A.F. - Astrophysical Observatory of Catania, Via S. Sofia 78, I-95123 Catania, Italy}
\altaffiltext{3}{Sydney Institute for Astronomy (SIfA), School of Physics, University of Sydney, NSW 2006, Australia}
\altaffiltext{4}{NASA-Ames Research Center, Moffett Field, CA 94035-0001, USA}
\altaffiltext{5}{Department of Physics and Astronomy, University of Victoria, PO Box 3055, Victoria, BC V8W 3P6, Canada}
\altaffiltext{6}{Instituut voor Sterrenkunde, K.U. Leuven, Celestijnenlaan 200D, 3001 Leuven, Belgium}
\altaffiltext{7}{LESIA - Observatoire de Paris, CNRS, Universit\'{e} Pierre et Marie Curie, Universit\'{e} Denis Diderot, 92195 Meudon cedex, France}
\altaffiltext{8}{Department of Astronomy, Yale University, P.O. Box 208101, New Haven, CT 06520-8101, USA}
\altaffiltext{9}{School of Physics and Astronomy, University of Birmingham, Edgbaston, Birmingham B15 2TT, UK}
\altaffiltext{10}{Kavli Institute for Theoretical Physics, Kohn Hall, University of California, Santa Barbara, CA 93106, USA}
\altaffiltext{11}{Danish AsteroSeismology Centre (DASC), Department of Physics and Astronomy, Aarhus University, DK-8000 Aarhus C, Denmark}
\altaffiltext{12}{Laboratoire AIM, CEA/DSM CNRS - Universit\'{e} Paris Diderot IRFU/SAp, 91191 Gif-sur-Yvette Cedex, France}
\altaffiltext{13}{Astronomical Institute ÒAnton PannekoekÓ, University of Amsterdam, Science Park 904, 1098 XH Amsterdam, The Netherlands}
\altaffiltext{14}{High Altitude Observatory, NCAR, P.O. Box 3000, Boulder, CO 80307, USA}
\altaffiltext{15}{Harvard-Smithsonian Center for Astrophysics, 60 Garden Street, Cambridge, MA, 02138, USA}
\altaffiltext{16}{Orbital Sciences Corporation/NASA Ames Research Center, Moffett Field, CA 94035}

\begin{abstract}
We studied solar-like oscillations in 115 red giants in the three open clusters NGC~6791, NGC~6811, and NGC~6819, based on photometric data covering more than 19 months with NASA's \textit{Kepler} space telescope. We present the asteroseismic diagrams of the asymptotic parameters $\dnua$, $\dnub$ and $\epsilon$, which show clear correlation with fundamental stellar parameters such as mass and radius. When the stellar populations from the clusters are compared, we see evidence for a difference in mass of the red giant branch stars, and possibly a difference in structure of the red clump stars, from our measurements of the small separations $\dnua$ and $\dnub$. Ensemble \'{e}chelle diagrams and upper limits to the linewidths of $\ell = 0$ modes as a function of $\Dnu$ of the clusters NGC~6791 and NGC~6819 are also shown, together with the correlation between the $\ell =0$ ridge width and the $\teff$ of the stars. Lastly, we distinguish between red giant branch and red clump stars through the measurement of the period spacing of mixed dipole modes in 53 stars among all the three clusters to verify the stellar classification from the color-magnitude diagram. These seismic results also allow us to identify a number of special cases, including evolved blue stragglers and binaries, as well as stars in late He-core burning phases, which can be potentially interesting targets for detailed theoretical modeling.
\end{abstract}

\keywords{open clusters and associations: individual (NGC~6791, NGC~6811, NGC~6819) -- stars: oscillations -- stars: evolution -- techniques: photometric}

\section{Introduction}
\label{sec:intro}
Many recent studies of solar-like oscillations of red giant stars have been focused on large ensembles of stars, made possible by the flood of high quality photometric data provided by the space missions CoRoT \citep[e.g.][]{DeRidder09,Kallinger10CoRoT,Mosser11universal} and \textit{Kepler} \citep[e.g.][]{Borucki10,Koch10,Bedding10redgiant}. Particular attention has been given to the three open clusters NGC~6791, NGC~6811, and  NGC~6819 in the \textit{Kepler} field \citep{Stello10cluster,Basu11,Hekker11cluster,Stello11amplitude,Stello11membership,Miglio11massloss}, due to the well-known advantage of cluster stars sharing common properties, which allows for more stringent investigations of stellar evolution theory. 

Among the highlights in recent results relevant for our study are the measurements of the small frequency separations $\dnua$, $\dnub$ and of the dimensionless term $\epsilon$, their correlation with the large frequency separation \citep[see][for previous results on field red giant stars]{Bedding10redgiant,Huber10redgiant,K12} and their dependence on stellar mass \citep{Montalban10redgiant,K12}. Also, the results on the ensemble \'{e}chelle diagrams have allowed for the investigation of ensemble properties of the modes, including the measurement of the mean small spacing $\dnuc$ \citep{Bedding10redgiant,Huber10redgiant} and the linewidths of the dipole modes and their correlation to fundamental stellar properties \citep[e.g. see][]{Chaplin09temp,Baudin11temp,App11temp,Belkacem2012}, which are important for the comprehension of the physics responsible for the excitation and damping of solar-like oscillations. Finally, the period spacing analysis for the investigation of the evolutionary stage of red giants \citep{Bedding11gmode,Mosser11redgiant} now allows us to distinguish between He-core burning red giants and those only burning hydrogen in a shell.

Here, we study 115 red giants belonging to the above mentioned clusters,  continuously observed for 19 months by the NASA \textit{Kepler Mission} \citep[see][for details on the data pipeline and acquisition, for a general introduction to the asteroseismic program, and for a description of the Kepler Input Catalog, respectively]{Jenkins10KAI,Gilliland10,Brown11}. Our study is made along the same lines as those described by \cite{Huber10redgiant}, who analyzed the first 4.5 months of \textit{Kepler} observations of field red giants. In contrast to \cite{Huber10redgiant}, our cluster red giants have the great advantage of providing more homogeneous samples because age, metallicity and mass \citep[see][]{Basu11,Miglio11massloss}, are about the same. In particular, NGC 6791 is a very old, \citep[$\sim 8.3$ Gyr,][]{Brogaard12}, and metal rich \citep[$\feh = 0.29 \pm 0.03$ (random) $\pm 0.07$ (systematic),][]{Brogaard11} open cluster, with average masses $M_\mathrm{RGB} = 1.20 \pm 0.01 \, M_\sun$ \citep{Basu11} and  $M_\mathrm{RC} = 1.15 \pm 0.03 \, M_\sun$ \citep{Miglio11massloss} for red giant branch (RGB) and red clump (RC) stars respectively \citep[see also][for recent results from eclipsing binaries]{Brogaard12}. NGC 6819 is a middle aged \citep[2$-$2.4 Gyr,][]{Basu11} open cluster, with solar metallicity \citep[$\feh = 0.09 \pm 0.03$,][]{Bragaglia01}, and average masses $M_\mathrm{RGB} = 1.68 \pm 0.03 M_\sun$ and $M_\mathrm{RC} = 1.65 \pm 0.04$ for RGB and RC stars respectively. The third open cluster, NGC 6811, is characterized by a young \citep[$0.7 \pm 0.1$ Gyr,][]{Glush99} and possibly solar metallicity star population (suggested by two independent spectroscopic investigations by Bruntt et al., in prep, and Molenda-\.Zakowicz et al., in prep.), where a small number of RC stars has been observed, showing an average mass $M_\mathrm{RGB} = 2.35 \pm 0.04 \, M_\sun$ \citep{Stello11amplitude,Stello11membership}. The temperature estimates for both NGC 6791 and NGC 6819 were derived by \cite{Hekker11cluster}. In particular, they used color-temperature calibrations by \cite{Ramirez05} and $JHK$ photometry from the 2MASS catalog \citep{2MASS}, which is available for all the stars of the sample. V magnitudes are taken from \cite{Stetson03} for NGC~6791 and from \cite{Hole09}, in order to derive temperatures based on the $( V - K)$ color. The adopted reddenings are $E(B-V) = 0.16 \pm 0.02$ for NGC~6791 \citep{Brogaard11} and $E(B-V) = 0.15$ for NGC~6819 \citep{Bragaglia01}. Lastly, \cite{Basu11} estimated DMs for both NGC~6791 and NGC~6819 by adopting the extinction $A_V = 3.1 E(B-V)$, which yielded to $(m-M)_0 = 13.11 \pm 0.06$ and $(m-M)_0 = 11.85 \pm 0.05$, respectively.

After briefly introducing the parameters involved in our study in Section~\ref{sec:asymp}, we describe the code developed for this work in Section~\ref{sec:data}, which concerns the analysis of the average p-mode structure in the power spectrum. In Section~\ref{sec:result} we show the resulting asteroseismic ensemble diagrams and the linewidths of radial modes as a function of fundamental stellar parameters, while Section~\ref{sec:gmode} presents the analysis of the period spacing of mixed dipole modes following the approach of \cite{Bedding11gmode}. Finally, we conclude in Section~\ref{sec:disc}.

\begin{figure}
\begin{center}
\includegraphics*[height=192px]{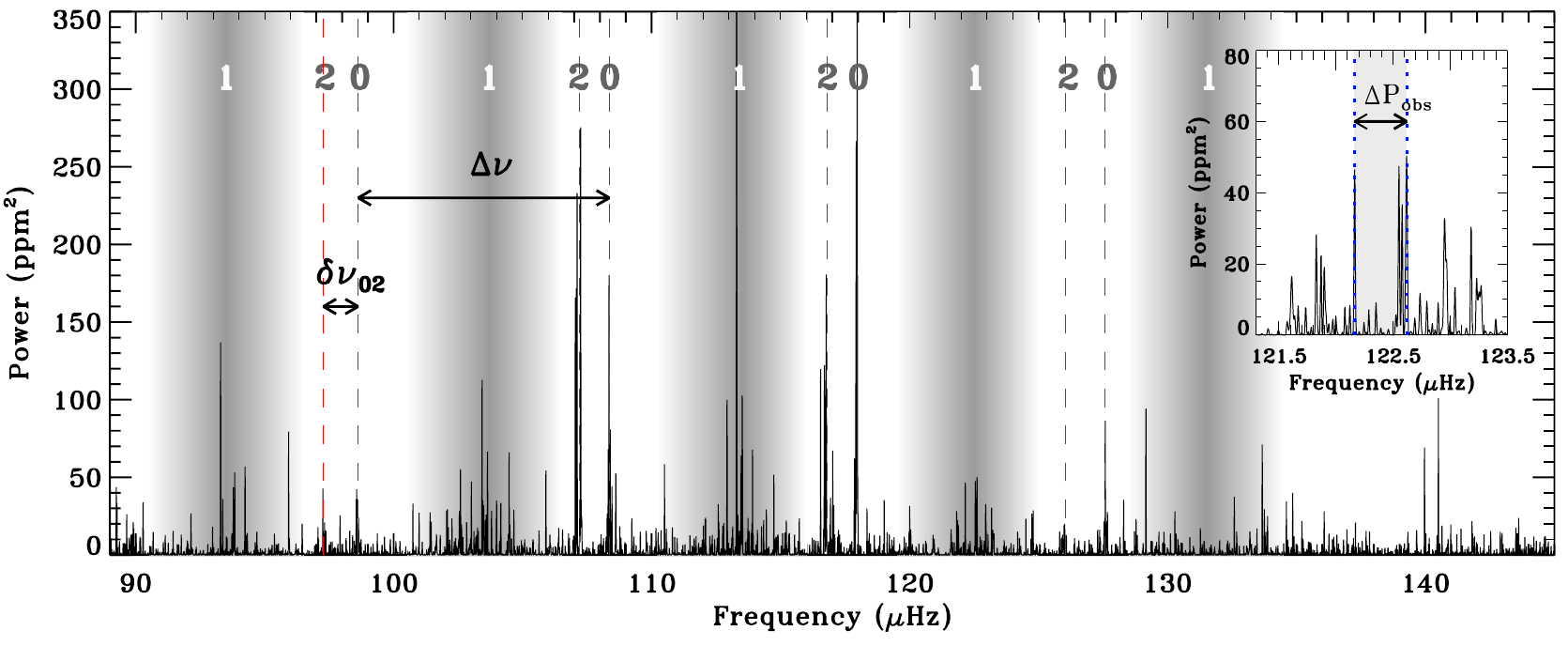}  
\caption{Power spectrum of KIC 2436593, a typical low luminosity RGB star belonging to NGC 6791. Mode identification for some of the peaks is shown. Shaded regions in gray indicate mixed $\ell = 1$ modes. $\Dnu$ and $\dnua$ are also marked. The inset shows the detail of one of the $\ell = 1$ shaded regions, where an indication of the observed period spacing, $\dP$, is shown.}
\label{fig:power}
\end{center}
\end{figure}

\section{Asymptotic parameters}
\label{sec:asymp}
Before proceeding with the description of the data analysis, it is helpful to introduce the physical quantities that we study in this work. As known from the asymptotic theory of solar-like oscillations, acoustic standing waves (also known as p modes) with low angular degrees, $\ell$, and high radial orders, $n$, show regular frequency spacings according to the relation \citep{Vandakurov68,Tassoul80,Gough86} 
\begin{equation}
\nu_{n \ell} \simeq \Dnu \left( n + \frac{\ell}{2} + \epsilon \right) - \delta\nu_\mathrm{0\ell} \, .
\label{eq:tassoul}
\end{equation}
Here 
\begin{equation}
\Dnu = \left( 2 \int^R_0 \frac{dr}{c(r)}\right)^{-1}
\end{equation}
is known as the large frequency separation, which scales roughly as the square root of the mean stellar density (Ulrich 1986), while $\epsilon$ is a phase shift sensitive to the properties of the near surface layers of the star \citep{CD92}. The term $\delta\nu_\mathrm{0\ell}$ is known as the small frequency separation and for $\ell = 1, 2$ and $3$ it is defined as
\begin{equation}
\dnua = \nu_\mathrm{n,0} - \nu_\mathrm{n-1,2},
\end{equation}
\begin{equation}
\dnub = \frac{1}{2} \left( \nu_\mathrm{n,0} + \nu_\mathrm{n+1,0} \right) - \nu_\mathrm{n,1},
\end{equation}
\begin{equation}
\delta\nu_\mathrm{03} = \frac{1}{2} \left( \nu_\mathrm{n,0} + \nu_\mathrm{n+1,0} \right) - \nu_\mathrm{n,3}.
\end{equation}
The small frequency separations are related to the sound speed gradient in the stellar core, hence to the mean molecular weight, which increases as the star evolves. Mixed modes occur as the frequencies of the g modes in the core and the p modes in the envelope become similar during the subgiant and red giant phase. As the star evolves, its mixed modes will undergo avoided crossings causing so called mode bumping, which broadens the ridges in the \'{e}chelle diagram \citep[e.g. see][]{Stello11nonradial}. As argued by \cite{Bedding10redgiant} and \cite{Huber10redgiant} the small spacing $\delta\nu_\mathrm{03}$ is therefore preferred in red giants over the more conventional $\delta\nu_\mathrm{13}$.
We note that mode bumping mostly affects the dipole modes as they penetrate deeper into the star, and hence couple more strongly to the g modes in the core \citep{Dupret09}. The dipole modes are therefore sensitive to the core properties of the star, which allows us to determine which red giants burn helium or not (see Section~\ref{sec:gmode}).

Figure~\ref{fig:power} shows a typical power spectrum of a low luminosity RGB star, KIC 2436593, observed in NGC 6791. The mode identification for some of the modes is shown, together with the indication of regions containing mixed $\ell = 1$ modes, represented by the gray-shaded strips. The large separation $\Dnu$ and the small spacing $\dnua$ are indicated as well. The inset shows a zoom-in of one of the gray-shaded strips, where the observed period spacing of the dipole modes, $\dP$ is marked.

\section{Observations and data analysis}
\label{sec:data}
The photometric time series of the 115 red giants used in this work were obtained in \textit{Kepler}'s long cadence mode \citep[$\Delta t \sim$ $30$ min,][]{Jenkins10} between 2009 May 13 and 2010 December 22. This corresponds to the observing quarters 1$-$7, providing a total of almost 18,000 data points per star \citep[see][for details on the detrending of the data]{Garcia11}. We followed the approach described by \cite{Stello11membership} for merging the quarters, and we discarded the stars that they classified as seismic non-members in their study. We also note that, according to their classification based on the color-magnitude diagram (CMD), the cluster stars in NGC~6811 are all He-burning stars, with one star appearing to be in a late He-core burning phase towards the asymptotic giant branch (AGB).

For the present study we developed the Asymptotic Analysis of Red Giants (AARG) code, with the purpose of deriving asymptotic parameters for p modes and observed period spacings for mixed modes in red giant stars. AARG performs a multi-step analysis in a semi-interactive way, allowing the user to follow the results at each step and make any necessary corrections. We calculated background-corrected power spectra and measured $\Delta \nu$ using the SYD pipeline \citep{Huber09pipeline}. As a check we compared $\Dnu$ values with those derived using other methods \citep{Mosser09pipe,Mathur10pipe,Hekker10pipe,Kallinger10} and found good agreement. We focus first on the analysis of p modes, which represents the main part of the work, leaving the discussion of period spacings to Section~\ref{sec:gmode}.

\begin{figure}
\begin{center}
\includegraphics*[height=192px]{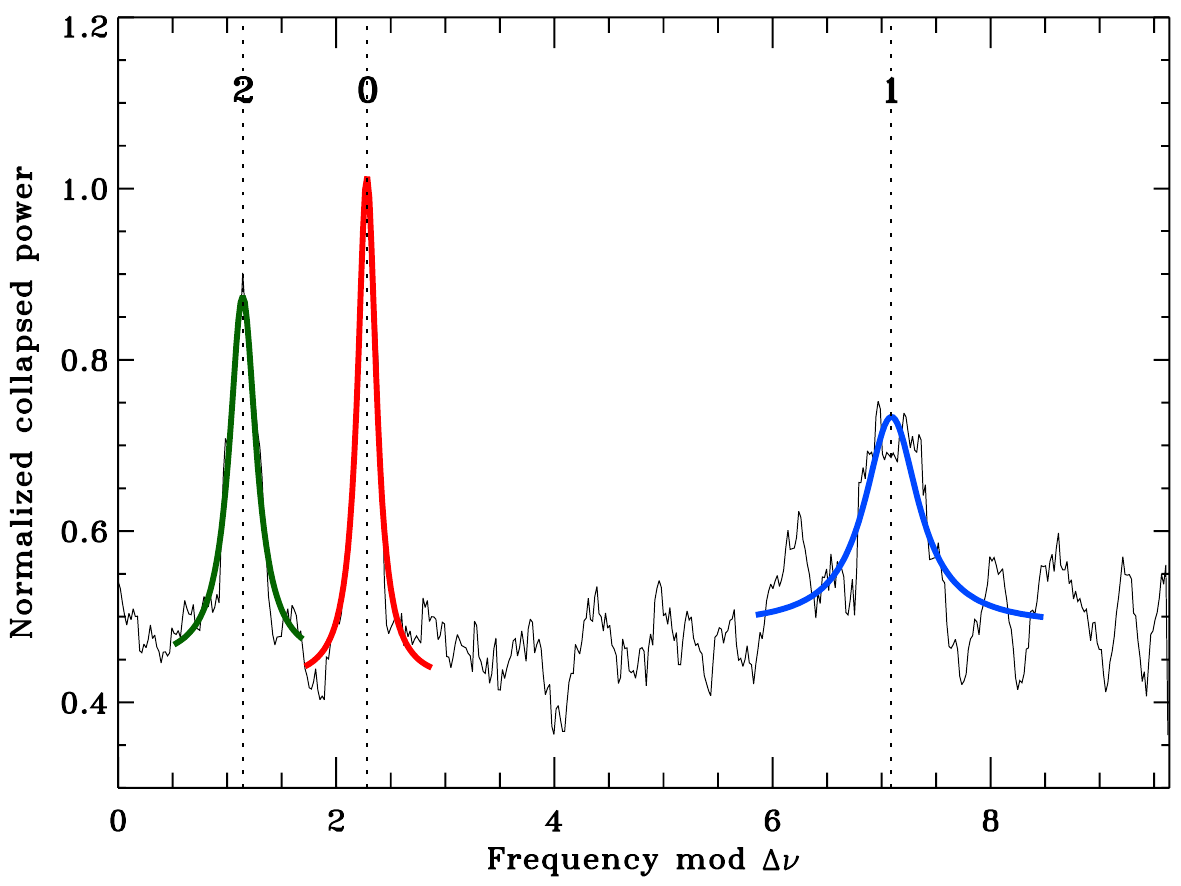}  
\caption{Collapsed \'{e}chelle diagram of KIC 2436593. The identification of the ridges $\ell = 0, 1$, and $2$ is shown, together with their Lorentzian fits (red, blue, and green solid lines, respectively).}
\label{fig:clps}
\end{center}
\end{figure}

The analysis of p modes, performed for each star, was done in three steps: (i) collapse the \'{e}chelle diagram using the measured $\Delta \nu$; (ii) identify the centroids $\nu_0,\nu_1,\nu_2$ of the $\ell = 0,1,2$ ridges by fitting three Lorentzian profiles to the collapsed \'{e}chelle diagram, which gives the small spacings $\delta\nu_\mathrm{02}$ and $\delta\nu_\mathrm{01}$, and $\epsilon$ \citep[see the next paragraph and][]{Huber10redgiant}; and (iii) simulate 500 power spectra by perturbing the observed power spectrum of the star according to a $\chi^2$ statistics with 2-degrees of freedom, perform the first two steps of the analysis for each simulation in order to derive a new set of asymptotic parameters, and evaluate their uncertainties by computing a robust rms of the results. Figure~\ref{fig:clps} shows an example of a collapsed \'{e}chelle diagram obtained with the AARG code. The centroids of the ridges $\ell = 0, 1$, and $2$ are marked by dotted lines, while the Lorentzian profiles used to fit the different ridges are shown with thick solid lines (red, blue, and green, respectively).

We now describe step (ii) in slightly more detail. AARG requires an initial guess of $\epsilon$ for each star, given by manually marking the position of the $\ell = 0$ peak in the collapsed \'{e}chelle diagram. This is followed by a Lorentzian fit to obtain the centroid, $\nu_0$. According to the values shown by \cite{Huber10redgiant} and \cite{White11}, we expected to have $0.5 < \epsilon < 1.5$ for red giants with $\Dnu < 15$ \muhz. We therefore either added or subtracted 1 to the measured $\epsilon$ to ensure it would fall within this range \citep[see also][for further discussion of the position of the centroids and the $\epsilon$ diagrams]{Mosser11universal,White11calc}.

Next, AARG makes a first guess for the centroid $\nu_2$ of the $\ell = 2$ peak by adopting an empirical relation $\nu_0 - \nu_2 \equiv \dnua = c \Dnu$, where $c$ is small. Although the relation has a slight mass dependence \citep[see][]{Montalban10redgiant}, using a fixed value of $c = 0.123$ offered a reliable first guess for every star in our sample \citep[note that our value is very close to the one measured by][]{Bedding10redgiant}. As for $\nu_0$, a Lorentzian fit centered on the first guess for the $\ell = 2$ ridge position provides the final value of the centroid $\nu_2$, and hence also $\dnua$. The search for the $\ell = 1$ peak is performed automatically by finding the maximum in the regions of the collapsed \'{e}chelle diagram laying outside the $\ell = 0,2$ peaks. A third Lorentzian fit is then performed, providing the centroid $\nu_1$, which gives $\dnub = \nu_0 + \Dnu / 2 - \nu_1$ according to the convention by \cite{Bedding10redgiant}. For a few stars ($\sim 10$) our method did not perform well. This was mainly caused by partly overlapping $\ell = 0, 2$ peaks (especially in NGC 6811 because of the higher mass of its stars) and strongly affected $\ell = 1$ peaks due to mixed modes. These stars were manually analyzed afterwards. We could successfully derive the asymptotic parameters of p modes for a total of 115 stars: 60 for NGC~6791, 5 for NGC~6811, and 50 for NGC~6819.

\section{Results}
\label{sec:result}
\subsection{$\epsilon$ diagram}
\label{sec:epsi}
\begin{figure}
\begin{center}
\includegraphics*[height=360px]{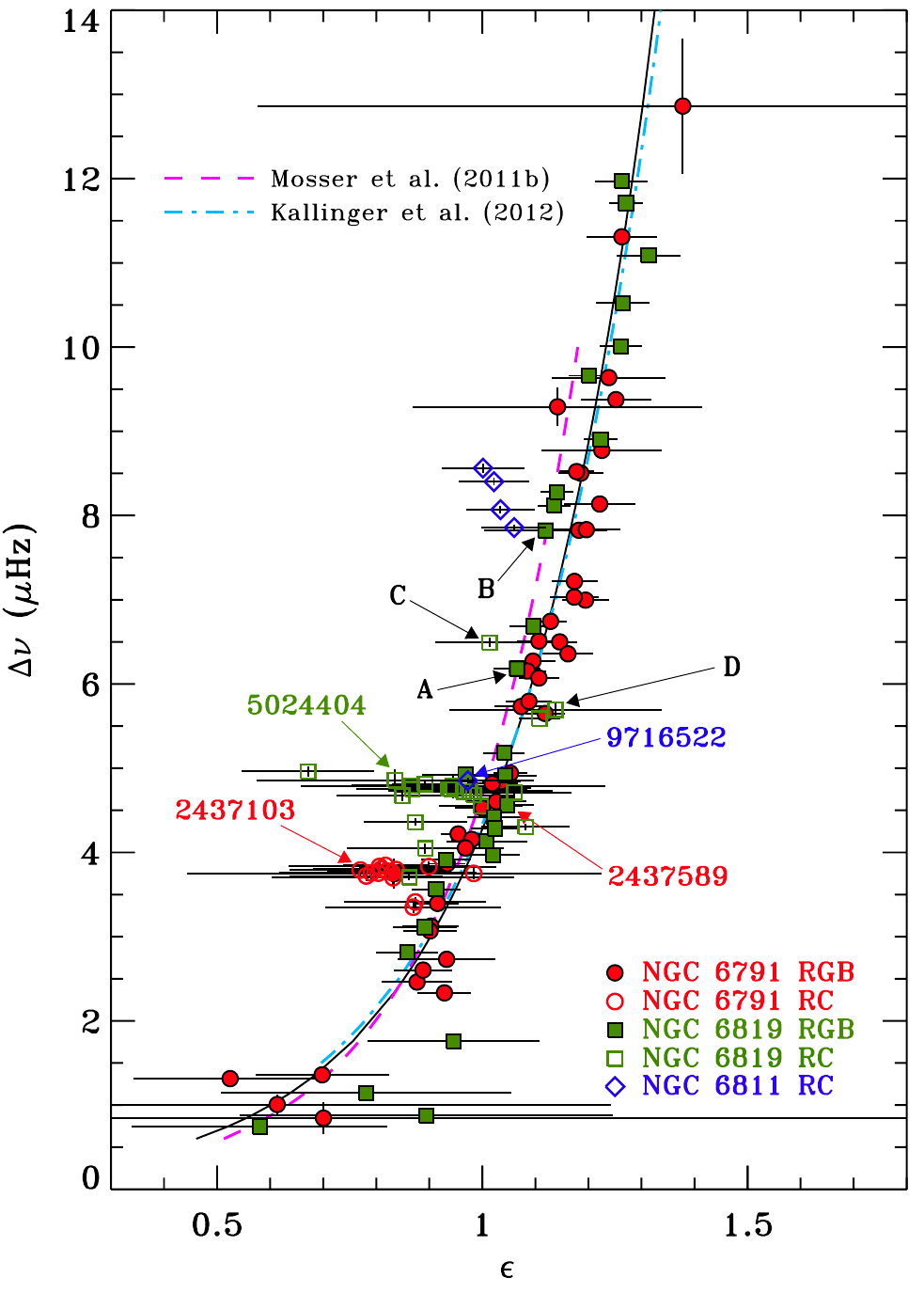}  
\caption{$\epsilon$ diagram for the clusters NGC~6791 (red circles), NGC~6811 (blue diamonds), and NGC~6819 (green squares). Open symbols represent RC stars while filled symbols are RGB stars. 1-$\sigma$ uncertainties are displayed for both quantities. A fit to the RGB stars of all the clusters using Equation~(\ref{eq:mosser}) is added (solid black line), as well as the one from \cite{Mosser11universal} (dashed purple line) and \cite{K12} (dot-dashed cyan line). Stars marked with labels and arrows represent special cases that are discussed in Section~\ref{sec:dpobs}.}
\label{fig:epsi}
\end{center}
\end{figure}

The $\epsilon$ term of Equation~(\ref{eq:tassoul}) was shown to be highly correlated with $\Dnu$ for red giant stars \citep{Huber10redgiant,Mosser11universal}.  The $\epsilon$ diagram is shown in Figure~\ref{fig:epsi} for the clusters NGC~6791, NGC~6811, and NGC~6819, where 1-$\sigma$ error bars were derived by means of Equation~(\ref{eq:tassoul}). We note that the RC stars (identified from the CMDs by \cite{Stello11membership} but adjusted for a few stars based on our analysis of the period spacing, presented in Section~\ref{sec:gmode}) form distinct groups with slightly lower $\epsilon$ than the RGB stars at $\Dnu \simeq 3.7 \, \mu$Hz for NGC~6791, $\Dnu \simeq 8 \, \mu$Hz for NGC~6811, and $\Dnu \simeq 4.8 \, \mu$Hz for NGC~6819. In particular, we measured a weighted average of $\epsilon$ for clump stars and RGB stars in the same $\Dnu$ range of RC stars, and found them to be $\langle \epsilon^\mathrm{RC} \rangle = 0.829 \pm 0.031$ and $\langle \epsilon^\mathrm{RGB} \rangle = 0.915 \pm 0.039$ for NGC 6791, and $\langle \epsilon^\mathrm{RC} \rangle = 0.970 \pm 0.018$ and $\langle \epsilon^\mathrm{RGB} \rangle = 1.015 \pm 0.017$ for NGC 6819. In both cases, $\langle \epsilon^\mathrm{RC} \rangle$ appears to be significantly different from $\langle \epsilon^\mathrm{RGB} \rangle$. This is in good agreement with \cite{Bedding11gmode} and \cite{K12}. Although a lower mass of RC stars can result in lower $\Delta \nu$, and hence lower $\epsilon$, one should note that this effect alone cannot explain the observed difference in $\epsilon$ between RC and RGB stars of similar $\Dnu$ \citep[see][for a detailed study about the mass difference between the RGB and RC stars]{Miglio11massloss}. The difference in evolutionary state also needs to be taken into account to fully explain the observed difference in phase shift \citep{K12}.

A least-squares fit to the RGB stars of the clusters was computed, using the log-relation
\begin{equation}
\epsilon = A + B \log \Dnu \, ,
\label{eq:mosser}
\end{equation}
adopted by \cite{Mosser11universal}. Since the fits computed to the RGB stars of NGC~6791 and NGC~6819 are not significantly different, we give the result for all the RGB stars in our sample, providing single values for the coefficients A and B. The result is shown as a solid black line in Figure~\ref{fig:epsi}, where $A = 0.601 \pm 0.025$ and $B = 0.632 \pm 0.032$. The fit from \cite{Mosser11universal}, who used a five-month data set, is added for comparison and plotted in the $\Delta \nu$ range $\left[ 0.6, 10 \right] \, \mu$Hz covered in that study (dashed purple line). The fit by \cite{K12}, based on more than 900 field red giants observed by \textit{Kepler} for about 600 days, is almost indistinguishable from ours (dot-dashed cyan line). We also tested a power-law form of the $\epsilon$-$\Dnu$ relation and found the $\chi^2$ to be very similar to that derived from Equation~(\ref{eq:mosser}). The log-relation was finally chosen to allow a direct comparison with the results by \cite{Mosser11universal} and \cite{K12}.

Lastly, we note that the uncertainties on $\Dnu$, and hence on $\epsilon$, become quite large for values of $\Dnu$ below $< 2 \, \mu$Hz due to the limited frequency resolution and small number of orders observed. For the star with the highest $\Dnu$, the large uncertainty is caused by the low S/N level, due to its low oscillation amplitude.

\subsection{C-D diagrams}
\label{sec:cddiag}
\begin{figure*}
\begin{center}
\includegraphics*[height=9cm]{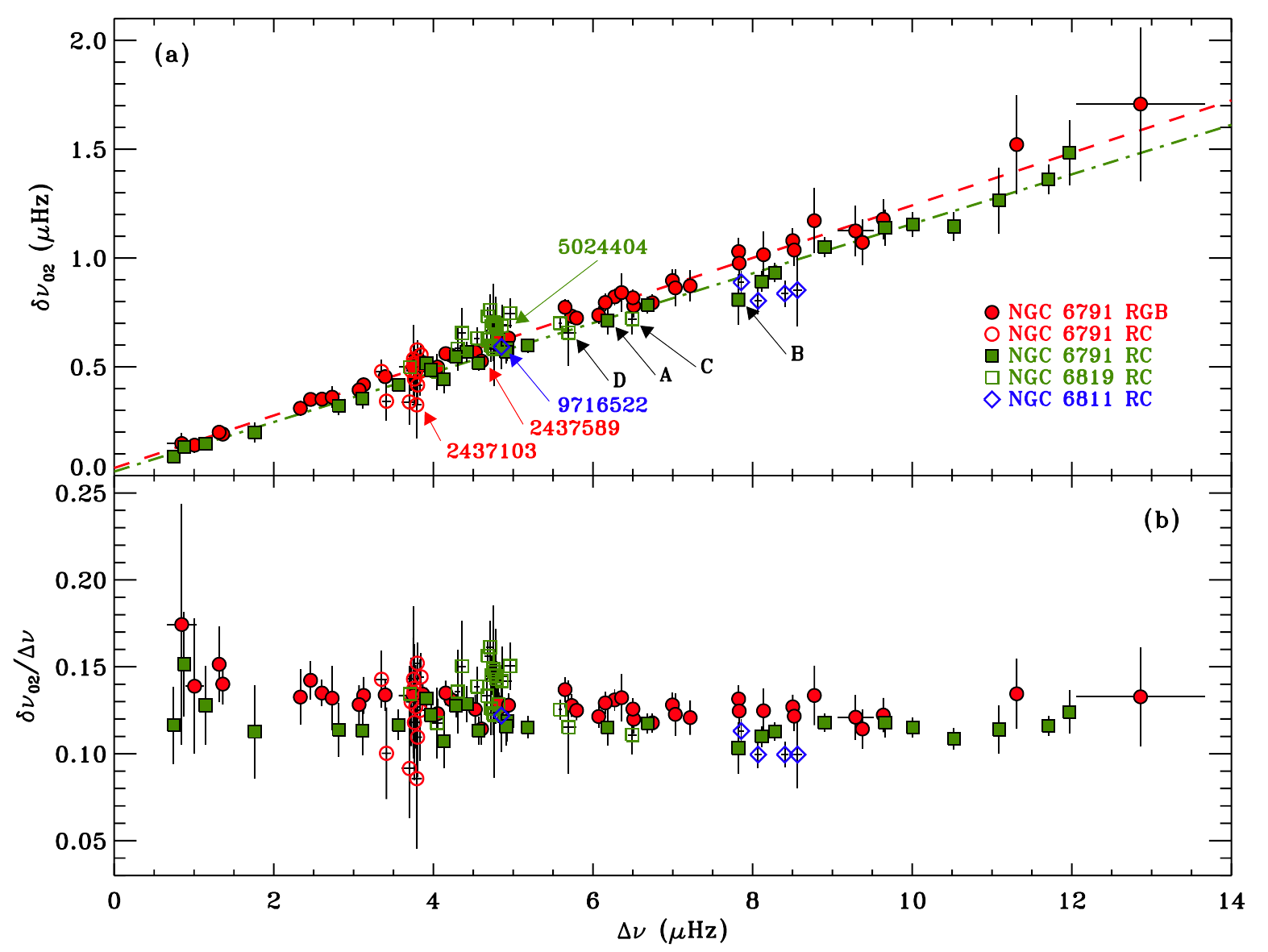} 
\includegraphics*[height=9cm]{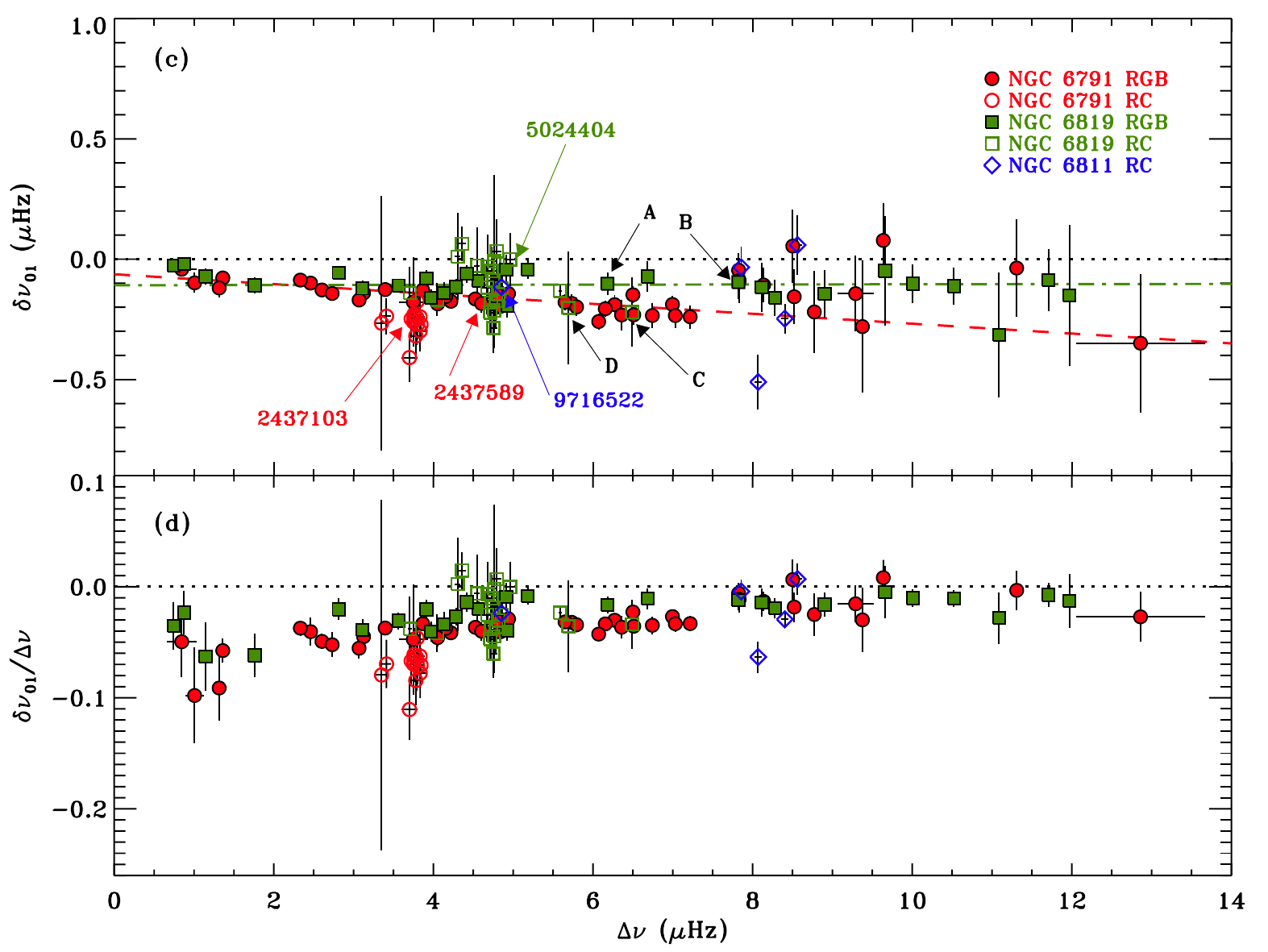} 
\caption{\textbf{(a)}, \textbf{(c)}: C-D diagrams of the small spacings $\dnua$ and $\dnub$  for the clusters NGC~6791 (red circles), NGC~6811 (blue diamonds), and NGC~6819 (green squares). Open symbols represent RC stars while filled symbols are RGB stars. Error bars show 1-$\sigma$ uncertainties. The linear fits to the RGB stars are shown for both NGC~6791 (dashed red line) and NGC~6819 (dot-dashed green line). Stars marked with labels represent special cases that are discussed in Section~\ref{sec:dpobs}. \textbf{(b)}, \textbf{(d)}: modified C-D diagrams of  the ratios $\dnua / \Dnu$ and $\dnub / \Dnu$ with the same notation adopted for the upper panels.}
\label{fig:cddiag}
\end{center}
\end{figure*}

In the C-D diagram one plots the small spacing $\dnua$ versus the large spacing $\Dnu$ \citep{CD84}, which for MS stars enables one to discriminate stars of different age and mass. A new version of the C-D diagram proposed by \cite{Mazumdar05} and by \cite{Montalban10redgiant} for MS and RGB stars respectively, is constructed by considering $\dnub$ instead of $\dnua$ \citep[see also][]{White11calc}. It has been shown that for red giants the C-D diagrams can not be used to investigate age \citep{White11} but that it is still useful to discriminate mass \citep[e.g. see][]{Bedding10redgiant,Huber10redgiant,Montalban10redgiant,K12}. The results for both $\dnua$ and $\dnub$ are shown in Figures~\ref{fig:cddiag}(a) and (c), for the three clusters. As before, open symbols represent RC stars while filled symbols are RGB stars.
As a first approximation, we represent the relation between $\dnua$ and $\Dnu$ by the linear relation $\dnua = a_\mathrm{02} + b_\mathrm{02} \Dnu$, which we fitted with a standard least-squares method to the RGB stars. The results are shown in Fig.~\ref{fig:cddiag} with a dashed red line for NGC~6791 and a dot-dashed green line for~NGC 6819. Their  equations are given by 
\begin{equation}
\delta\nu^\mathrm{(6791)}_\mathrm{02} = \left( 0.121 \pm 0.003 \right) \Dnu + \left( 0.035 \pm 0.012 \right) \, \text{\muhz} \, ,
\label{eq:6791}
\end{equation}
and
\begin{equation}
\delta\nu^\mathrm{(6819)}_\mathrm{02} = \left( 0.114 \pm 0.003 \right) \Dnu + \left( 0.019 \pm 0.012 \right) \, \text{\muhz} \, .
\label{eq:6819}
\end{equation}
Only error bars on $\dnua$ were considered for the fits, but the results obtained by including uncertainties on both quantities were indistinguishable from the ones presented here. The coefficients of the $\dnua$-$\Dnu$ relation estimated from our fit agree within a few percent with those derived by \cite{K12} for field stars. 

The typical mass for an RGB star, $M_\mathrm{RGB}$, is expected to be different for each cluster but about the same within a given cluster. For cluster RGB stars we therefore have a much tighter constraint on the stellar mass than for field stars. \cite{Basu11} found $M_\mathrm{6791} = 1.20 \pm 0.01 \, M_{\sun}$ and $M_\mathrm{6819} = 1.68 \pm 0.03 \, M_{\sun}$ as the averages for the RGB stars, which were derived using grids of stellar models that incorporated scaling relations for $\nu_\mathrm{max}$ and $\Dnu$. We refer to \cite{Miglio11massloss} for further discussion about the mass estimates for these stars. Following the theoretical work by \cite{Montalban10redgiant}, who showed that $\dnua$ depends on mass, we relate the difference in the slopes, $b_\mathrm{02}$, in Equations~(\ref{eq:6791}) and~(\ref{eq:6819}) to the difference in $M_\mathrm{RGB}$. Assuming the linear relation
\begin{equation}
b_\mathrm{02} = \alpha_\mathrm{02} + \beta_\mathrm{02} \left( \frac{M_\mathrm{RGB}}{M_\sun} \right) \, ,
\label{eq:mass_lr}
\end{equation}
we obtain $\alpha_\mathrm{02} = 0.138 \pm 0.012$ and $\beta_\mathrm{02} = -0.014 \pm 0.008$ by solving the system of two equations (one for each cluster). 

As done for $\dnua$, we fitted a linear relation $\dnub = a_\mathrm{01} + b_\mathrm{01} \Dnu$ to the RGB stars of NGC~6791 and NGC~6819. The linear trends are shown in Figure~\ref{fig:cddiag}(c) with the same notation as Figure~\ref{fig:cddiag}(a), and the results are
\begin{equation}
\delta\nu^\mathrm{(6791)}_\mathrm{01} = - \left( 0.021 \pm 0.003 \right) \Dnu - \left( 0.063 \pm 0.011 \right) \, \text{\muhz}
\label{eq:6791d01}
\end{equation}
and
\begin{equation}
\delta\nu^\mathrm{(6819)}_\mathrm{01} = \left( 0.000 \pm 0.003 \right) \Dnu - \left( 0.109 \pm 0.012 \right) \, \text{\muhz} \, .
\label{eq:6819d01}
\end{equation}
Once again, the uncertainties are quite similar for the two clusters. Like $\dnua$, we also see a mass dependence on $\dnub$ for the RGB stars. But unlike $\dnua$, the trend appears to go in the opposite direction, with higher $\dnub$ for higher masses. This is in qualitative agreement with the theoretical results by \cite{Montalban10redgiant}, whose Figure~5(b) shows a slight increase in $\dnub$ for increasing mass along the RGB.  As for $\dnua$, we relate the slopes $b_\mathrm{01}$ in Equations~(\ref{eq:6791d01}) and ~(\ref{eq:6819d01}) to $M_\mathrm{RGB}$, assuming the linear relation 
\begin{equation}
b_\mathrm{01} = \alpha_\mathrm{01} + \beta_\mathrm{01} \left( \frac{M_\mathrm{RGB}}{M_\sun} \right) \, ,
\label{eq:mass_lr2}
\end{equation}
and find $\alpha_\mathrm{01} = -0.073 \pm 0.012$ and $\beta_\mathrm{01} = 0.044 \pm 0.008$. We find that $| \beta_\mathrm{01} |\simeq 3 | \beta_\mathrm{02}|$, hence it appears that $\dnub$ is more sensitive to mass than $\dnua$ by about a factor of three. But at this stage we would caution overinterpretation of this result
as further theoretical investigations are required to fully understand how $\dnub$ depends on the fundamental parameters and internal structure of red giants (Section~\ref{sec:unexp}).

Figures~\ref{fig:cddiag}(b) and (d) show the so-called modified C-D diagrams which plot the relative ratios $\dnua / \Dnu$ and $\dnub / \Dnu$. The reason for considering the ratio $\delta\nu_\mathrm{0\ell} / \Dnu$ is that models show it to be less sensitive to surface layer effects \cite[e.g. see][]{White11calc} and that the small spacings $\delta\nu_\mathrm{0\ell}$ approximatively scale with $\Dnu$. Our results appear to be in agreement with previous results on red giants \citep[][]{Bedding10redgiant,Huber10redgiant,Mosser11universal,K12} and with the theoretical studies by \cite{Montalban10redgiant}.

\subsection{The small spacings of red clump stars}
\label{sec:unexp}
It is interesting to compare the average small spacings for the RC stars relative to the RGB stars in each cluster.
In the following we denote this quantity $\Delta \langle \delta \nu_\mathrm{0 \ell} \rangle \equiv \langle \delta \nu^\mathrm{RC}_\mathrm{0 \ell} \rangle -  \langle \delta \nu^\mathrm{RGB}_{0 \ell} \rangle$. It is evident in all four panels of Figure~\ref{fig:cddiag}, but slightly more so in Figures~\ref{fig:cddiag}(b) and (d), that the RC stars on average show different small spacings than RGB stars of similar $\Dnu$. We will first discuss $\Delta \langle \dnua \rangle$. 

For NGC 6819, $\Delta \langle \dnua \rangle = 0.112
\pm 0.016 \, \mu$Hz, while for NGC 6791 we have $\Delta \langle \dnua \rangle = 0.012 \pm 0.021 \, \mu$Hz. 
Given the relation between $\dnua$ and mass for RGB stars (Section~\ref{sec:cddiag}), one might speculate that a similar relation would exist for RC stars. However, we note that $\Delta \langle \dnua \rangle$ for NGC 6819 is about twice as large as the difference in $\dnua$ between the two RGB populations in NGC 6791 and NGC 6819 at a similar $\Dnu$. Hence, if we applied the relation in Equation~(\ref{eq:mass_lr}) to the RC stars, we would find that the RC stars in NGC~6819 have a mass of about $0.7 \, M_\sun$ (corresponding to a mass loss of about $1 \, M_\sun$), in stark disagreement with the results by \cite{Miglio11massloss}, who found $\Delta \langle M \rangle = -0.03 \pm 0.04$. 
Hence, there is certainly something else dominating the different values of $\Delta \langle \dnua \rangle$ we see for the two clusters.

Turning our attention to the other small spacing, we have $\Delta \langle \dnub \rangle = 0.004 \pm 0.025 \, \mu$Hz for NGC~6819 and $\Delta \langle \dnub \rangle = -0.113 \pm 0.020 \, \mu$Hz for NGC~6791. Hence we see that $\Delta \langle \dnub \rangle_\mathrm{6791} < \Delta \langle \dnub \rangle_\mathrm{6819}$, which was also the case for $\Delta \langle \dnua \rangle$. We recall that the mass dependencies of $\dnua$ and $\dnub$ were opposite for the RGB, both in observations (Section~\ref{sec:cddiag}) and models \citep{Montalban10redgiant}. Hence, we would also expect an opposite trend for $\Delta \langle \dnub \rangle$ ($\Delta \langle \dnub \rangle_\mathrm{6791} > \Delta \langle \dnub \rangle_\mathrm{6819}$). The fact that we do not observe this is further evidence that a simple relation with mass alone cannot explain the observed differences in small spacings between RC and RGB stars.

A possible explanation is the internal structural changes of the stars that occur during the He-flash phase \citep{Bildsten12} between the tip of the RGB and the RC. These changes could be significantly different for stars of different masses ($M_\mathrm{6791} = 1.20 \pm 0.01 \, M_\sun$ and $M_\mathrm{6819} = 1.68 \pm 0.03 \, M_\sun$, \cite{Basu11}), composition ([Fe/H]$_\mathrm{6791} = 0.29 \pm 0.03$ (random) $\pm 0.07$ (systematic), \cite{Brogaard11}; [Fe/H]$_\mathrm{6819} = 0.09 \pm 0.03$, \cite{Bragaglia01}), and rotation rates \citep{Meibom11}. Further investigation requires modeling of both the RGB and RC stars in these clusters. 

The dependence of $\dnub$ on stellar properties was investigated by \cite{Montalban10redgiant} using stellar models covering $0.7-2.3 \, M_\sun$ on the RGB and $2.5-5.0 \, M_\sun$ in the He-core burning phase. They found that small values of $\dnub$ were predominantly seen among RGB models, and we would therefore expect the RC stars to show larger $\dnub$ on average, which is in contrary to what we observe for NGC 6791. However, we note that all the He-core burning models in the \cite{Montalban10redgiant} sample were more massive than the stars in the two clusters considered here, and a direct comparison is therefore not possible. The physical cause of a lower value of $\dnub$ was not firmly established by \cite{Montalban10redgiant}, but they argued that there was a tendency for low $\dnub$ values in models where the inner turning point of the $\ell=1$ modes was well inside the convective envelope, corresponding to stars with deep convective envelopes. Clearly, these issues deserve further study.

\begin{figure*}
\begin{center}
\includegraphics*[height=385px]{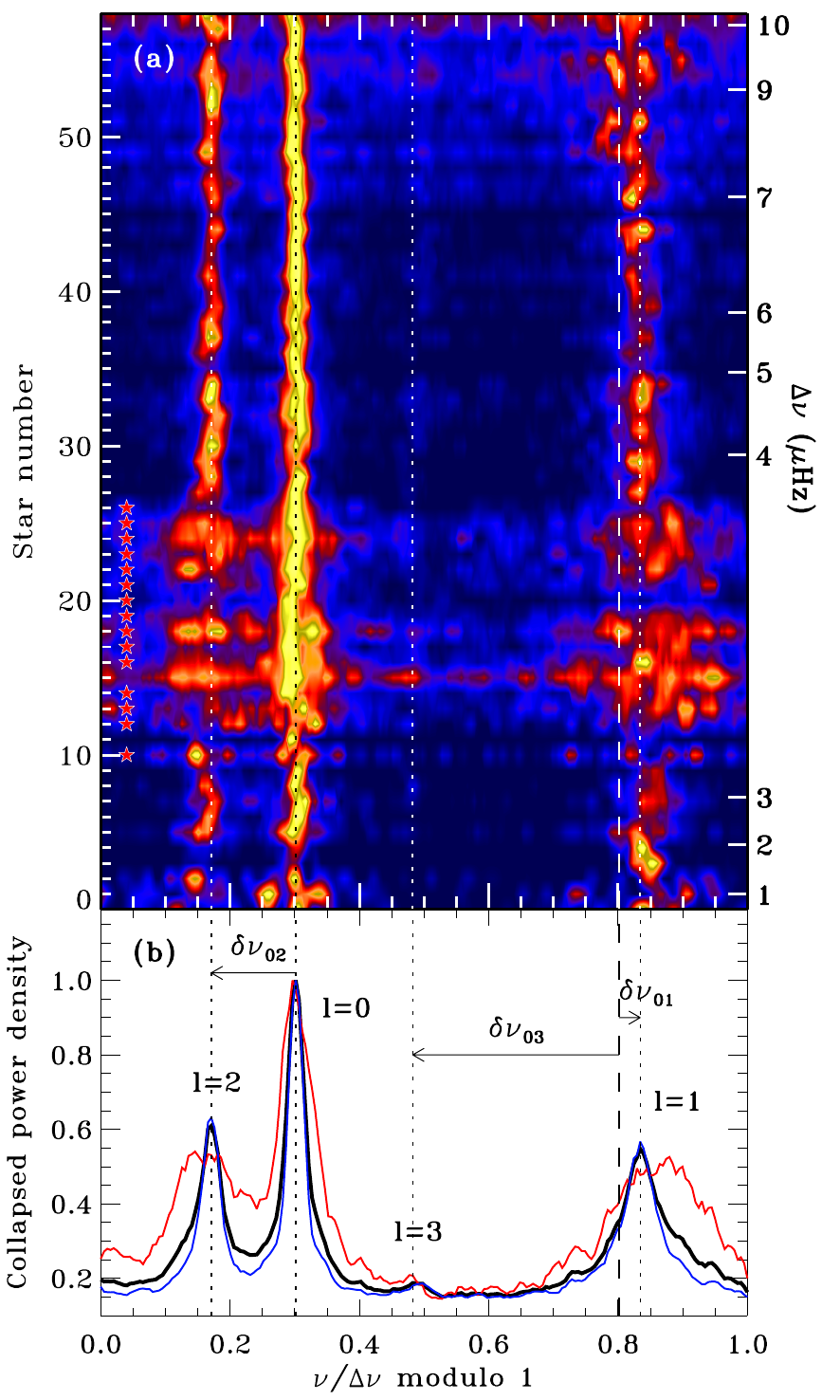} 
\includegraphics*[height=385px]{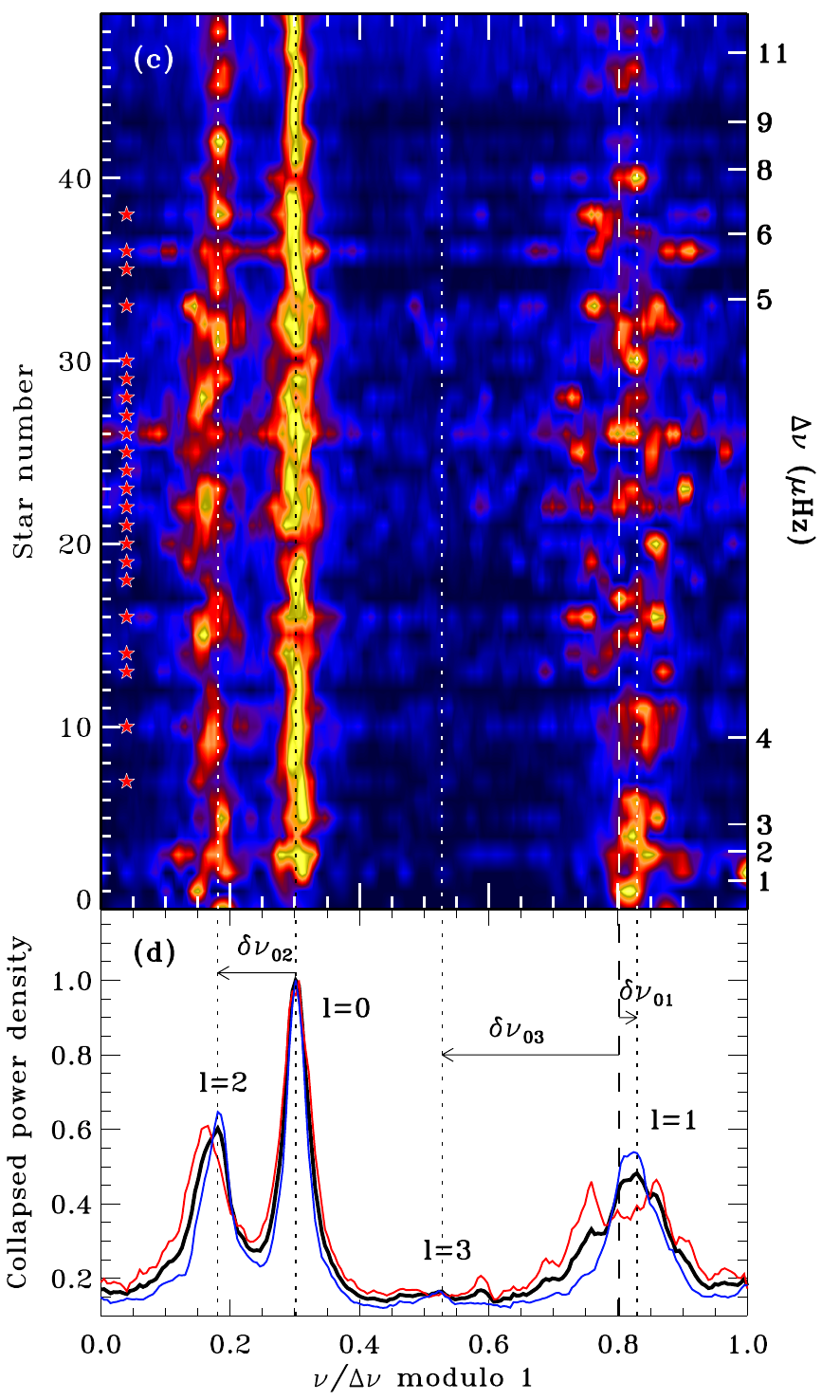}
\caption{\textbf{(a)}, \textbf{(c)}: ensemble  \'{e}chelle diagrams of the clusters NGC~6791 and NGC~6819, respectively, where $\ell = 0$ centroids were aligned by shifting the $\ell = 0$ ridge of each star to align with ($\nu / \Dnu$ mod $1) = 0.3$. The number of the stars, ordered by increasing $\Dnu$, is shown on the left coordinate, and the corresponding $\Dnu$ is shown on the right axis. Red star symbols mark the clump stars identified in the clusters. Note that each row corresponds to the collapsed \'{e}chelle of one star, normalized to unity. \textbf{(b)}, \textbf{(d)}: diagrams showing panels (a), (c) collapsed over the entire range of $\Dnu$ (thick black line) normalized to unity. Results for RC stars in red and RGB stars in blue are also shown. Ridge identifications and definitions of small separations are indicated. In both panels, the dotted lines represent the centroids of the $\ell = 0,1, 2$ and $3$ ridges, while the dashed line is the position of the midpoint of two adjacent $\ell = 0$ modes.}
\label{fig:ensemble}
\end{center}
\end{figure*}

\subsection{Ensemble \'{e}chelle diagrams}
\label{sec:ensemble}
Following \cite{Huber10redgiant}, we computed the so-called ensemble \'{e}chelle diagrams for both NGC~6791 and NGC~6819. When dealing with a large number of stars, ensemble \'{e}chelle diagrams are very helpful for studying the evolution of features such as ridge width and position. In particular, the measurement of the average position of the $\ell = 3$  ridge becomes possible also when low S/N in the power spectra does not allow one to make a clear detection of the corresponding peak in a single star. The results are shown in Figures~\ref{fig:ensemble}(a) and (c), where the stars are numbered by increasing $\Dnu$. Each row in the plot represents the collapsed \'{e}chelle diagram of a single star using the large separation adopted in the analysis, and shifted in order to have the $\ell = 0$ ridge fall on $(\nu / \Dnu$ mod $1) = 0.3$ \citep[see also the discussion by][Sec. 2]{Stello11nonradial}. The RC stars (red star symbol)  clearly show strong broadening of the $\ell = 1$ and 2 ridges. We see that even the $\ell = 0$ ridge appears broader for RC stars in both clusters.

The result of collapsing the ensemble \'{e}chelle over the entire sample of stars is shown in Figures~\ref{fig:ensemble}(b) and (d) (thick black line). Results for RC stars (red line) and RGB stars (blue line) are also plotted for both clusters. The presence of an $\ell = 3$ peak becomes evident for NGC~6791, while for NGC~6819 a hint of $\ell=3$ is visible only for  the RGB stars. For NGC~6791, the $\ell = 3$ hump seems to arise from several stars, particularly those with $\Dnu < 7 \, \mu$Hz, as visible from Figure~\ref{fig:ensemble}(a). The position of the marked $\ell = 3$ peaks of the two clusters, and hence their average small spacings $\dnuc$, are in agreement with the results of \cite{Bedding10redgiant,Huber10redgiant,Mosser11universal} and \cite{K12}. 

It is noticeable that the $\ell = 1,2$ ridges move away from the $\ell = 0$ ridge as the stars evolve from H-shell to He-core burning red giants, a result that was already discussed by \cite{Huber10redgiant}. We also note that the hump visible in Figure~\ref{fig:ensemble}(d), on the left slope of the $\ell = 1$ peak (red line), is caused by only two stars having strong peaks that occur at $(\nu / \Dnu$ mod $1) \simeq 0.7$ and this is therefore not an indication of a general feature. Referring to the effect on $\dnub$ discussed in Section~\ref{sec:unexp}, we notice that the $\ell = 1$ ridge of the RC stars of NGC~6791 (Figure~\ref{fig:ensemble}(a)) is shifted towards the right-hand side of the diagram, i.e towards lower values of the small spacing, while this shift is not apparent in NGC~6819 (Figure~\ref{fig:ensemble}(c)).

\begin{figure}
\begin{center}
\includegraphics*[height=165px]{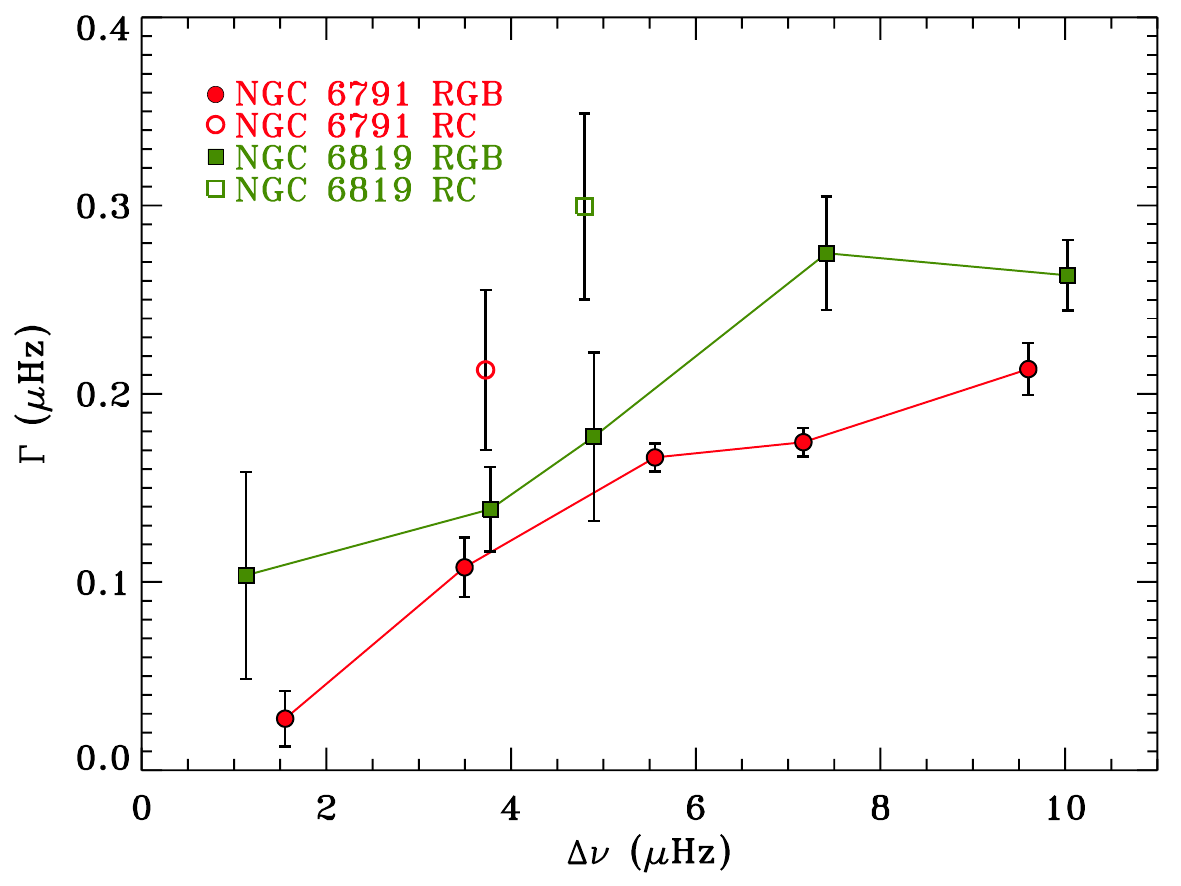}
\caption{FWHM of the $\ell = 0$ ridge as a function of $\Dnu$ for RGB stars in NGC~6791 (filled red circles) and NGC~6819 (filled green squares). Each point represents the average of values within a subset of stars with similar $\Dnu$. The error bars are the 1-$\sigma$ uncertainties on the mean for each subset. Open symbols at $\Dnu \simeq 3.7 \, \mu$Hz and $\Dnu \simeq 4.8 \, \mu$Hz represent the measurements for the subsets of RC stars.}
\label{fig:fwhm}
\end{center}
\end{figure}

\begin{figure}
\begin{center}
\includegraphics*[height=165px]{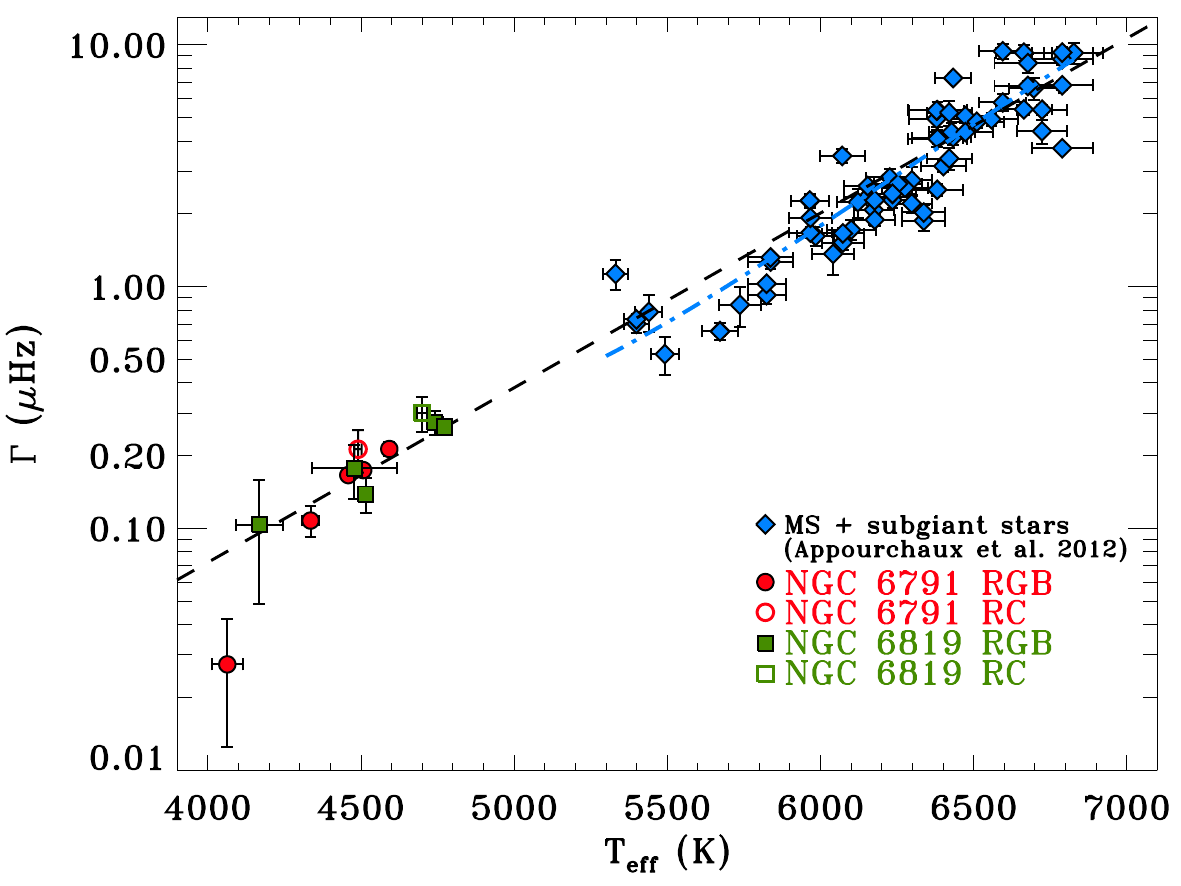}
\caption{FWHM of the $\ell = 0$ ridge plotted against $\teff$ for the stars of NGC~6791 (red circles) and NGC~6819 (green squares). Also shown are measured linewidths for MS and subgiant field stars (blue diamonds) from \cite{App11temp}. Each cluster point represents the same subset of stars plotted in Figure~\ref{fig:fwhm}. The error bars are the 1-$\sigma$ uncertainties on the mean for each subset. The fit to the MS and subgiant  stars taken from \cite{App11temp} is also shown (dot-dashed blue line). The dashed black line shows an exponential fit (Equation~(\ref{eq:teff})) to all stars.}
\label{fig:teff}
\end{center}
\end{figure}

\subsection{Mode linewidths}
\label{sec:fwhm}
Measuring the linewidths of p modes and studying how they correlate to the fundamental stellar properties has important consequences for the understanding of the damped nature of solar-like oscillations. In fact, the physics responsible for the damping mechanism that acts in the convective envelope of low mass stars is not yet fully understood \citep[e.g. see][]{Houdek99,Dupret09,Belkacem2012}.

In the present work we provide estimates of the linewidths of radial modes derived through the AARG code.
In particular, the widths of the ridges in the collapsed \'{e}chelle diagrams (Figures~\ref{fig:ensemble}(b) and (d))
give a rough estimate of the mode linewidths.
Figure~\ref{fig:fwhm} shows the FWHM for the $\ell = 0$ ridge from the
Lorentzian fit to the corresponding peak in the collapsed \'{e}chelle
diagram, for the RGB stars of both NGC~6791 and NGC~6819. Each point is the
average from a subset of stars sorted in bins of $\Dnu$, while the overlaid
error bars are 1-$\sigma$ uncertainties on the mean for each bin. The open
symbols at $\Dnu \simeq 3.7 \, \mu$Hz and $\Dnu \simeq 4.8 \, \mu$Hz show our
measurements for the RC stars. We see a clear increasing trend when moving
to higher $\Dnu$, a result that was already apparent from the
analysis by \cite{Huber10redgiant} of field red giants, despite of the
shorter data set that was available \citep[see also][who obtained a similar
 result by using a different method]{K12}. This increasing trend is also
visible in Figures~\ref{fig:ensemble}(a) and (c), in that the scaled width
$\nu / \Dnu$ of the $\ell = 0$ ridge at low $\Dnu$ is about the same of
that at high $\Dnu$. In Figure~\ref{fig:fwhm} we also notice a systematic
difference between the ridge widths of the two clusters, a feature that is
already visible from the collapsed \'{e}chelle diagrams of
Figures~\ref{fig:ensemble}(b) and (d). 

To see whether the difference in ridge width between the two clusters and between stars with different
$\Dnu$ arises from the difference in temperature of the stars as contemplated by \cite{Chaplin09temp,Baudin11temp,App11temp,Belkacem2012}, we plot our measurements of FWHM as a function of $\teff$ in a log-log scale in Figure~\ref{fig:teff}. This shows indeed that all the cluster stars
follow an almost common trend, which supports that the observed
difference in ridge width largely follows the difference in temperature. 
We also show the linewidth measurements of a sample of main sequence (MS) and
subgiant stars (blue diamonds) from \cite{App11temp}, where we have taken
temperatures from \cite{Casagrande06,Casagrande10}.
It is remarkable how well all the stars are aligned in Figure~\ref{fig:teff}. Note that our measure of the ridge width only
provides an upper limit to the `true' mode linewidths because of the slight
curvature of the ridges in the \'{e}chelle diagram.
The fit to the linewidths across all stars is represented by an exponential function
\begin{equation}
\Gamma = \Gamma_0 \, exp{\left( \frac{\teff - 5777 \, \text{K} }{T_0} \right)}  \, \text{\muhz} \, ,
\label{eq:teff}
\end{equation}
where $\Gamma_0 = 1.39 \pm 0.10 \, \mu$Hz and $T_0 = 601 \pm 3\,$K (dashed black line).
A detailed study using linewidths found by direct mode fitting (peak bagging) of MS and red giant stars
\cite[e.g.,][]{Chaplin09temp,Baudin11temp,App11temp} goes beyond the scope
of this work. The power law fit with a background component proposed by \cite{App11temp} is here added for comparison in its range of validity ($5300 \,$K$-$$6800\,$K, dot-dashed blue line).
However, we can conclude that our measurements, combined with \textit{Kepler} 
results on MS and subgiant stars, follow a single exponential trend with temperature.

\section{Mixed modes}
\label{sec:gmode}
Mixed modes have the great advantage of being sensitive to the core structure, while at the same time being observable at the surface. They were recently used as a way to successfully distinguish between RC and RGB stars \citep{Bedding11gmode,Mosser11redgiant}. Although their amplitude is lower than of pure p modes \citep{Mosser12}, long datasets enable us to identify many of them due to their long lifetimes \citep{Dupret09}. Even in cluster red giants, which are generally fainter than the \textit{Kepler} field stars, we can detect many mixed $\ell = 1$ modes in the best cases. The main features of mixed modes relevant for the analysis presented in this work are discussed in Section~\ref{sec:ps}, while our results on their period spacings are described in Section~\ref{sec:dpobs}. 

\subsection{Period spacings of mixed dipole modes}
\label{sec:ps}
While p modes are equally spaced in frequency, pure g modes are approximately equally spaced in period, following the asymptotic relation \citep{Tassoul80,CD11redgiant}
\begin{equation}
\Pi_{n \ell} = \frac{\Delta \Pi_\mathrm{0}}{\sqrt{\ell \left( \ell + 1\right)}} \left(n + \alpha \right) \, .
\label{eq:gmode}
\end{equation}
Here, $n$ and $\ell$ are the radial order and the angular degree of the mode, $\alpha$ is a small constant and $\Delta \Pi_\mathrm{0}$ is the period spacing, expressed as
\begin{equation}
\Delta \Pi_\mathrm{0} = 2\pi^2 \left( \int \frac{N}{r} dr \right)^{-1} \, .
\end{equation}
The integral is taken over the cavity in which the g modes propagate, and $N$ is the Brunt-V\"{a}is\"{a}l\"{a} frequency. From Equation~(\ref{eq:gmode}), the period spacing of dipole g modes is given by 
\begin{equation}
\Delta P_\mathrm{g} = \Delta \Pi_\mathrm{0} / \sqrt{2} \, , 
\label{eq:pspacing}
\end{equation}
which appears the most interesting quantity to investigate because of the strong coupling between p and g modes for $\ell = 1$ \citep{Dupret09,Montalban10redgiant,Stello11nonradial,BeddingWS}.

However, in contrast to the large separation for p modes, the period spacing of pure g modes, $\Delta P_\mathrm{g}$, cannot always be directly measured in red giants because all the non-radial modes are mixed in the red giant phase \citep{CD11redgiant}. Nevertheless, from recent studies it seems to be possible to infer $\Pi_\mathrm{0}$ in some cases \citep[see][]{Bedding11gmode,Mosser12mixed}. Fortunately we can readily measure the period spacing of the mixed modes, $\Delta P_\mathrm{obs}$, which can serve as a proxy for $\Delta P_\mathrm{g}$.  $\dP$ is lower than $\Delta P_\mathrm{g}$ by about a factor of 0.6-0.8 \citep[e.g. see][]{Bedding11gmode,Mosser12mixed}.

\begin{figure*}
\begin{center}
\includegraphics*[height=8cm]{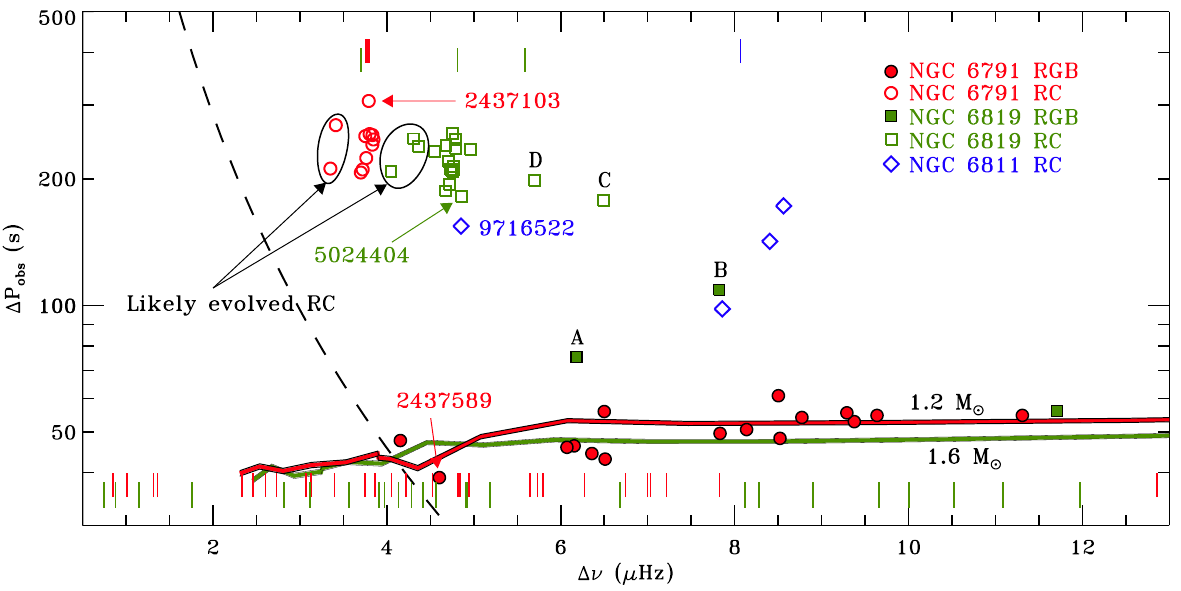}
\caption{Period spacings of the three clusters NGC~6791 (red circles), NGC~6811 (blue diamonds), and NGC~6819 (green squares). Open symbols represent RC stars, while filled symbols are RGB stars. Tracks for $1.2$ $M_\sun$ (thick red line) and $1.6$ $M_\sun$ (thick green line), and $Z = 0.017$, are shown \citep{White11calc}. Tick marks drawn at the top (RC) and bottom (RGB) refer to stars that could not be identified with our period spacing analysis, and are colored according to the notation adopted in the rest of the paper. The dashed black line represents the minimum period spacing one can measure with a 19 months-long time-series. Special cases discussed in Section~\ref{sec:dpobs} and listed in Table~\ref{tab:outliers} are marked.}
\label{fig:dpobs}
\end{center}
\end{figure*}

\subsection{$\dP$-$\Dnu$ diagram}
\label{sec:dpobs}
As mentioned in Sections~\ref{sec:intro} and~\ref{sec:data}, AARG measures period spacings using the approach used by \cite{Bedding11gmode}. As a first step, it modifies the power spectrum for each star by erasing the regions containing all the $\ell = 0, 2$ modes, whose positions come directly from our analysis of p modes  (see Section~ \ref{sec:data}). This new power spectrum shows only $\ell =1$ mixed modes, and possibly some low amplitude $\ell = 3$ modes, and is then expressed in period rather than frequency. The power spectrum of this power spectrum is then calculated, which is converted back into period. To obtain a first guess for the observed period spacing, a manual marking of the position of the excess of power is required. Finally, a Gaussian fit to the selected hump of power provides our measure of $\dP$.

We were able to measure $\dP$ for a total of 53 stars in our sample: 27 from NGC~6791, 4 from NGC~6811, and 22 from NGC~6819. The results are shown in Figure~\ref{fig:dpobs}, where RC and RGB stars are marked with open and filled symbols, respectively. The dashed black line is the limit set by the frequency resolution. Overlaid are theoretical tracks for $1.2 \, M_\sun$ and $1.6 \, M_\sun$ at near-solar metallicity (Z = 0.017), as calculated by \cite{White11calc}, which are representative of the RGB stars of NGC~6791 and NGC~6819, respectively. Using stellar models, we verified that changing the metallicity over the range spanned by the two clusters has no significant effect on $\Delta P_\mathrm{g}$ for RGB stars.

The tick marks at the top (RC) and bottom (RGB) represent stars for which the period spacing could not be clearly measured by our analysis, classified by \cite{Stello11membership} using the CMD. We see that the fraction of stars with measured period spacings is much higher for RC stars than for RGB stars even after taking into account the limit set by the frequency resolution. In particular, for NGC~6791 these fractions are $\sim$36\% (RGB) and $\sim$73\% (RC), while for NGC~6819 they are $\sim$10\% (RGB) and $\sim$86\% (RC). This could be explained by a weaker coupling between the p-mode and g-mode cavities for the RGB stars \citep{Dupret09}, which makes the resonances narrower in frequency, resulting in a smaller number of observable mixed modes.

\subsection{Discussion of special cases}
\label{sec:special}
The stars labeled from A to D (KIC 5112361, KIC 4937770, KIC 5024414 and KIC 5024476) are outliers in the $\dP$-$\Dnu$ diagram (Figure~\ref{fig:dpobs}), while KIC 2437103, KIC 2437589, and KIC 5024404 have period spacings that imply a different stage of evolution to the one based on the CMDs \citep{Stello11membership}. KIC 9716522 represents a star on its way up towards the AGB, as already noted by \cite{Stello11membership} and now supported by our measurement of its high period spacing in agreement with that of other He-burning stars (Figure~\ref{fig:dpobs}).

\begin{figure}
\begin{center}
\includegraphics*[height=5.8cm]{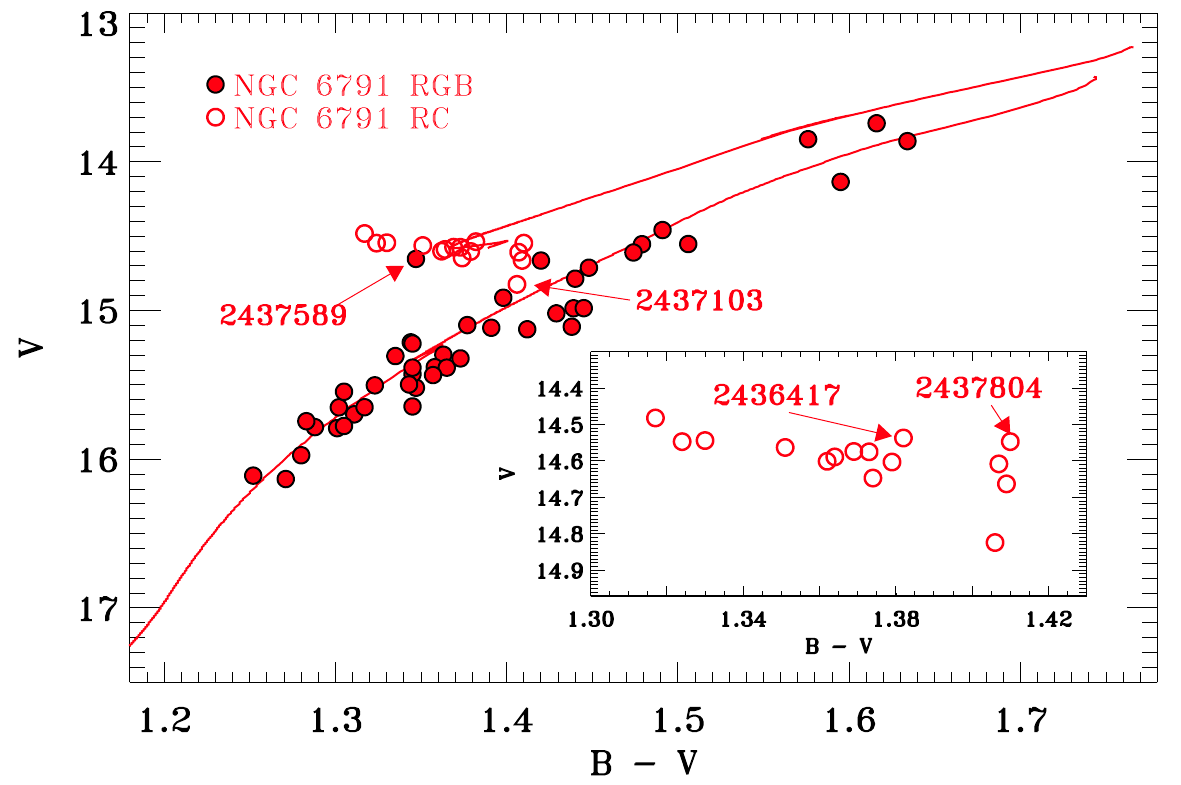}
\includegraphics*[height=5.8cm]{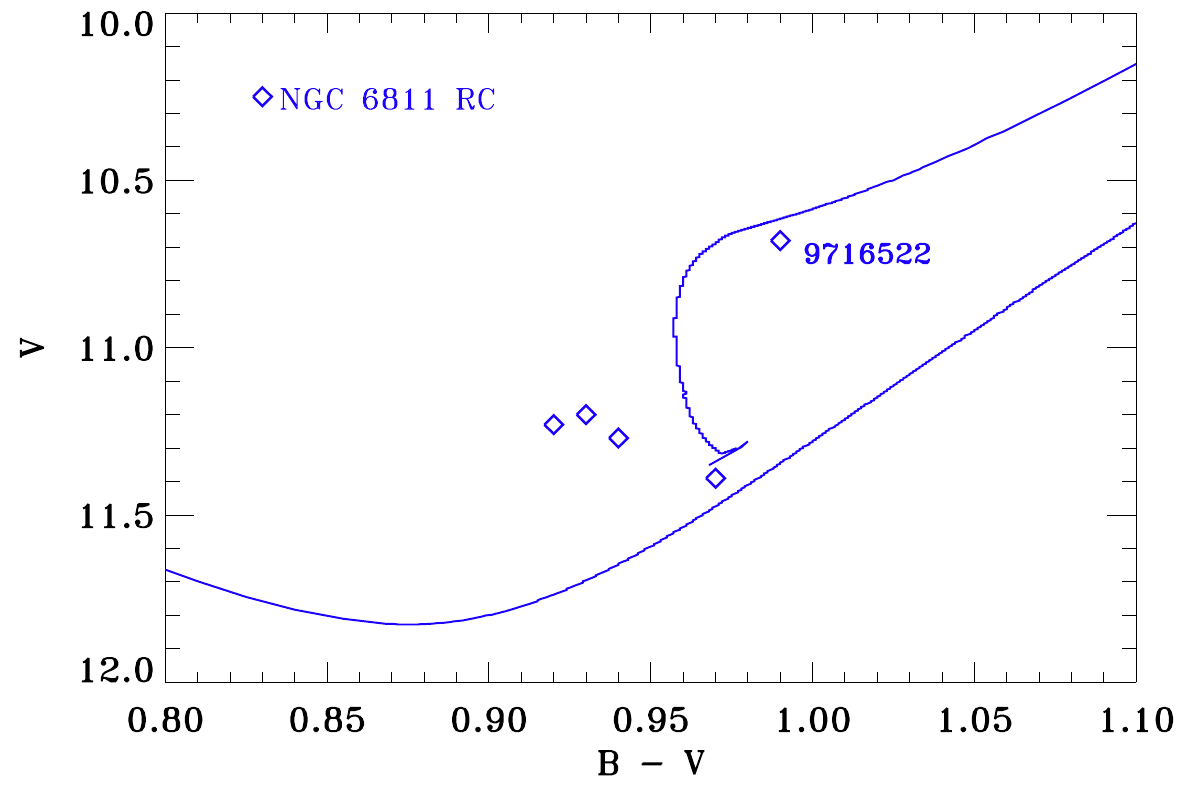}
\includegraphics*[height=5.8cm]{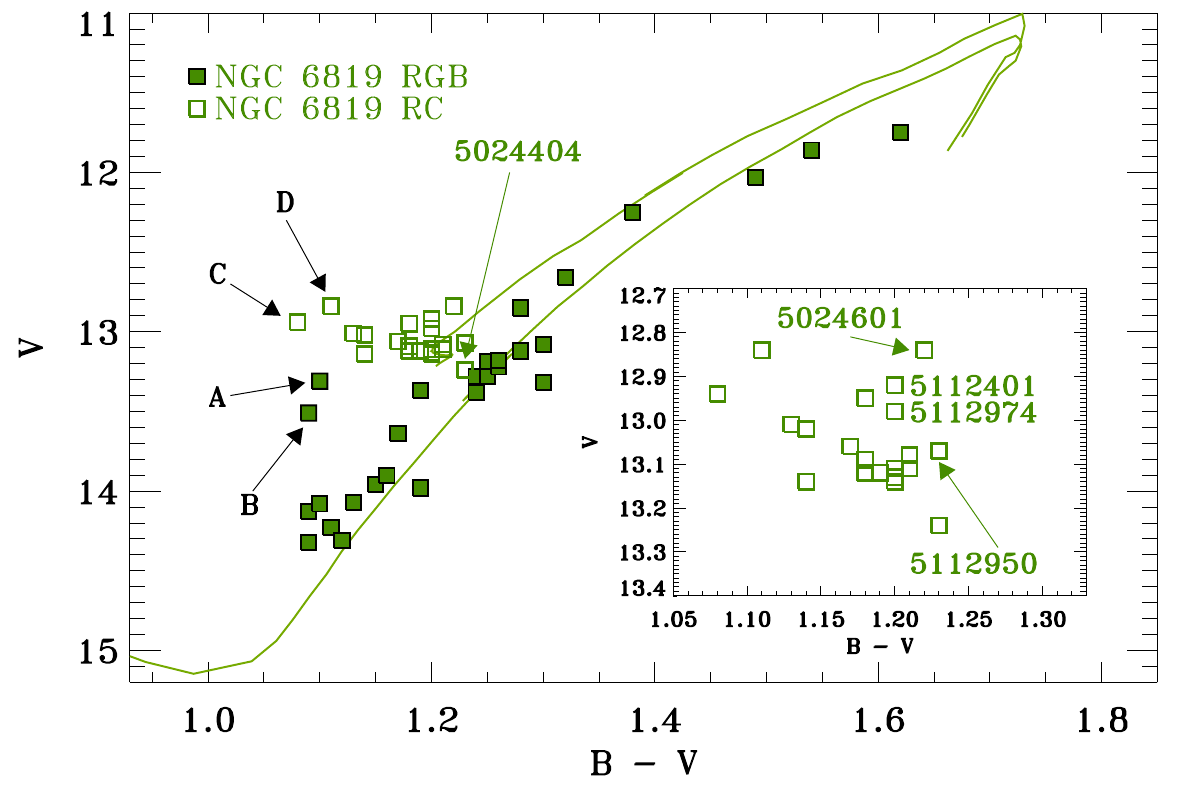}
\caption{CMDs of the clusters NGC~6791 (top panel), NGC~6811 (middle panel), and NGC 6819 (bottom panel) as derived by \cite{Stello11membership}. Both RC and RGB stars are shown, with open and filled symbols respectively, according to the classification obtained by the membership study of \cite{Stello11membership} and our analysis of period spacings. Stars marked with labels represent special stars discussed in Section~\ref{sec:dpobs} and listed in Table~\ref{tab:outliers}. Isochrones are shown for all the clusters \citep[solid lines, see][for details]{Stello11membership}.}
\label{fig:cmd}
\end{center}
\end{figure}

\begin{figure}
\begin{center}
\includegraphics*[height=6.5cm]{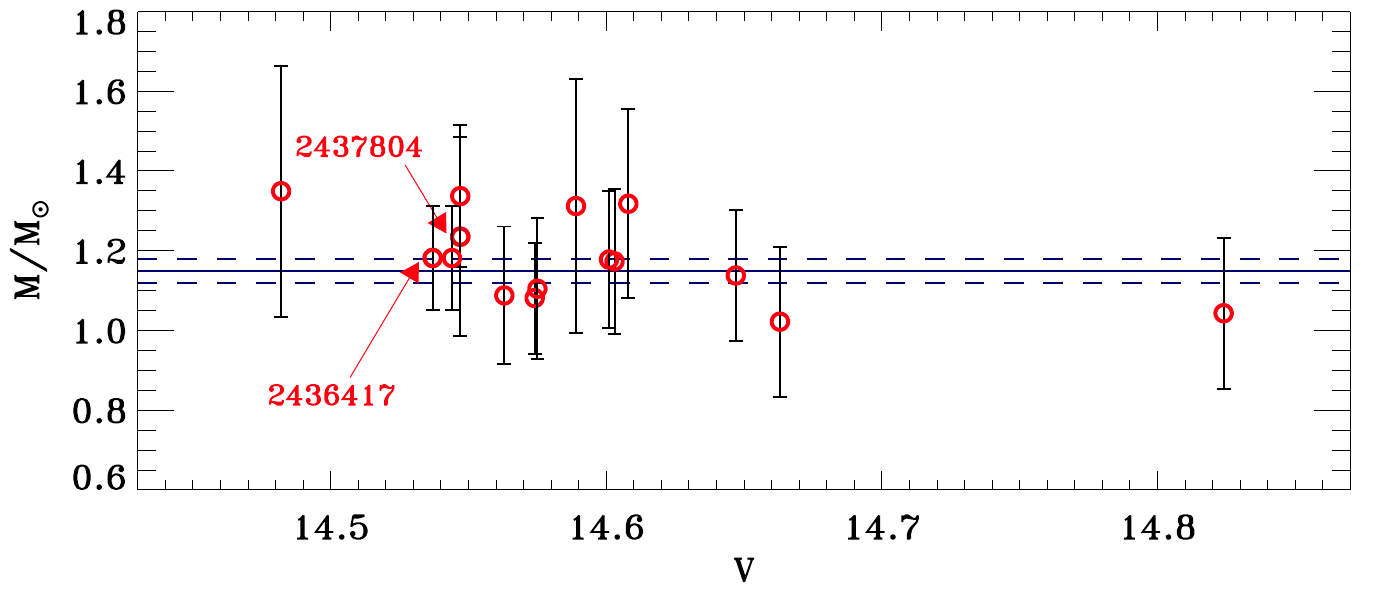}
\includegraphics*[height=6.5cm]{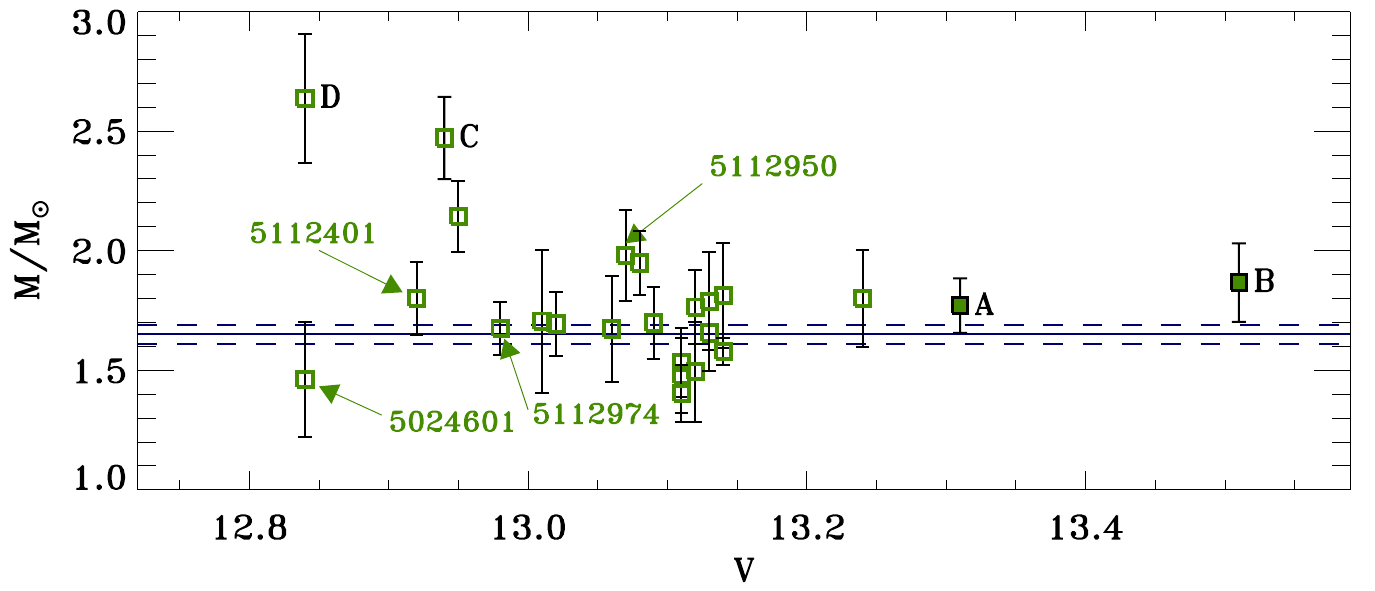}
\caption{Mass of stars near the RC of NGC~6791 (red circles) and NGC~6819 (green squares) with applied correction in the $\Dnu$ scaling of $2.7$ \% and $1.9$ \% respectively \citep[see][]{Miglio11massloss}. Error bars show $1-\sigma$ uncertainties derived according to Equation~(\ref{eq:scal_mass}). Outlier stars A$-$D and `likely evolved RC' stars discussed in Section~\ref{sec:dpobs} and listed in Table~\ref{tab:outliers} are marked. Blue lines represent the corrected mean masses of RC stars (solid) and their 1-$\sigma$ uncertainties (dashed) derived by \cite{Miglio11massloss}.}
\label{fig:mass}
\end{center}
\end{figure}

All these highlighted stars are also marked in Figures~\ref{fig:epsi} and \ref{fig:cddiag}, and in the CMDs of Figure~\ref{fig:cmd} \citep[as derived by][]{Stello11membership}. We also list these stars in Table~\ref{tab:outliers}, together with all their asteroseismic parameters derived in this work. To further support the discussion presented below, we derived the masses of all stars near the RC in the CMD, including the outliers A$-$D and the stars that we have marked as `likely evolved RC' (Figure~\ref{fig:dpobs}), which are also shown in Figure~\ref{fig:cmd} and listed in Table~\ref{tab:outliers}. To estimate the masses we use the scaling relation
\begin{equation}
\frac{M}{M_\sun} \simeq \left( \frac{\nu_\mathrm{max}}{\nu_\mathrm{max,\sun}} \right)^3 \left( \frac{\Dnu}{\Dnu_\sun} \right)^{-4} \left( \frac{\teff}{T_\mathrm{eff,\sun}} \right)^{3/2} 
\label{eq:scal_mass}
\end{equation}
where we adopted $\nu_\mathrm{max,\sun} = 3100 \, \mu$Hz, $\Dnu_\sun = 135 \, \mu$Hz and $T_\mathrm{eff,\sun} = 5777 \,$K \citep[e.g.][]{Miglio11massloss}. The result is shown in Figure~\ref{fig:mass}, with masses plotted against $V$ magnitude and 1-$\sigma$ error bars overlaid. Blue lines represent the mean masses of RC stars (solid) and their 1-$\sigma$ uncertainties (dashed), as derived by \cite{Miglio11massloss} by adopting Equation~(\ref{eq:scal_mass}). To provide corrected estimates of mass for clump stars, the $\Dnu$ scaling relation was corrected by 2.7~\% and 1.9~\% for NGC 6791 and NGC 6819, respectively, according to the study by \cite{Miglio11massloss}.

We first discuss possible causes for the outliers, A$-$D. All
four are potentially binary stars.  Three of them (A, C, and D) are listed
as binary stars in the radial velocity study by \cite{Hole09}, and the
fourth star (B) shows a low oscillation amplitude, which could be
indicative of a binary star, as argued by \cite{Stello11amplitude}.  
All four stars also appear relatively blue in the CMD (Figure~\ref{fig:cmd}). 
Stars A and B fall below the RC in the CMD and are in
line with the rest of the RGB stars in Figures~\ref{fig:epsi} and \ref{fig:cddiag}, suggesting that
they are RGB stars with no clear sign of an abnormal mass
(Figure~\ref{fig:cddiag}(a) and (b)). This is confirmed by our estimate of their masses 
according to Equation~(\ref{eq:scal_mass}) (Figure~\ref{fig:mass}), whose values are similar to 
the average mass of the RGB stars of NGC~6819 found by \cite{Basu11} (see Section~\ref{sec:cddiag}). 
Binarity seems like the most plausible explanation
for their $B-V$ colors being lower than the other RGB stars.  However, their power
spectra do not show oscillations from two components, and their higher-than-expected
$\dP$ is therefore difficult to explain. Perhaps it could come from a different core structure
of these stars caused by binary interaction.  We note that the stars do
not seem to be the result of a merger event, given their apparently `normal'
masses. In conclusion, stars A and B are most likely both binaries, with one
component on the RGB, whose seismic signal we detect, and a fainter less-evolved component. 
Stars C and D have luminosities typical to
that of the RC. Our measurement of $\dP$ suggests
that the stars indeed belong to the RC. The position of the stars in the
sequence of He-core burning stars going from low mass (low $\Dnu$) to high
mass (high $\Dnu$) spanned by the three clusters indicates that stars C and D
have higher masses than the other RC stars in NGC~6819 (Figure~\ref{fig:dpobs}). 
This is confirmed by our estimate of their masses (Figure~\ref{fig:mass}). 
Our conclusion that they are high-mass RC stars is in good agreement with \cite{RV98}, who
mention these stars along with others with this position in the CMD to be potential descendants of blue stragglers, meaning that
they experienced mass transfer and therefore have a component with a mass
significantly higher than the cluster's turn-off mass.  

Concerning the next three stars, our period spacing analysis shows that KIC 2437103 ($\dP = 306 \,$s) is an RC star, and KIC 2437589 ($\dP = 39 \,$s) is an RGB star, as argued by \cite{Miglio11massloss}, and KIC 5024404 ($\dP = 182 \,$s) is an RC star. It seems that KIC 2437589 is an evolved blue straggler in the RGB phase, as suggested by \cite{Brogaard12}. This would explain its unusual position in the CMD (top panel of Figure~\ref{fig:cmd}), and is also supported by a mass of about $1.7 \, M_\sun$, as derived from Equation~(\ref{eq:scal_mass}), greater than the mass of the other RGB stars of the cluster. 

Lastly, six stars (two in NGC~6791 and four in NGC~6819), were found to be possible candidates for RC stars that are starting to evolve towards the AGB. We list them as `likely evolved RC' in Table~\ref{tab:outliers}. Our suggestion arises from our measurement of their $\dP$, which corresponds to that of RC stars, and from their $\Dnu$, which is lower than that of the other RC stars. Their masses (Figure~\ref{fig:mass}) are similar to that of the average RC star which, in combination with their lower $\Dnu$, confirm that they have a radius significantly larger than the other RC stars.

\section{Summary and Conclusion}
\label{sec:disc}
To summarize and conclude on the main results of our analysis:
\begin{enumerate}
\item The fit of the $\epsilon$-$\Dnu$ relation to the RGB stars of our sample, computed using Equation~(\ref{eq:mosser}), is compatible with the result derived by \cite{Mosser11universal}, although it deviates slightly towards higher values of $\Dnu$, where our sample has more stars and benefits from longer observations. Our fit is almost indistinguishable from that by \cite{K12}, which was based on more than 900 field red giants observed for a similar length of time. Moreover, we tested a power-law form of the $\epsilon$-$\Dnu$ relation and found it to provide a very similar fit to that derived from Equation~(\ref{eq:mosser}). Lastly, the average $\epsilon$ of clump stars appears to be significantly different from that of their RGB counterparts for both NGC~6791 and NGC~6819, a result in agreement with previous findings on field RGs.

\item The linear fits to the $\dnua$-$\Dnu$ relation for the RGB stars of our sample, given by Equations~(\ref{eq:6791}) and ~(\ref{eq:6819}), appear to be compatible within a few percent with the results by \cite{Huber10redgiant} and \cite{K12} on field red giants. A direct measure of the mass-dependence for the small spacings $\dnua$ and $\dnub$ is derived for the first time for cluster stars. The result indicates that $\dnub$ is about three times more sensitive to a mass difference than $\dnua$. The mass-dependence for $\dnua$ is compatible with the results by \cite{K12} on field red giants. Furthermore, both $\dnua$ and $\dnub$ show dependence on mass that is qualitatively in agreement with theoretical studies of red giant stars by \cite{Montalban10redgiant}.

\item It is notable that the RC stars of NGC~6791 behave differently from those of NGC~6819 for both $\dnua$ and $\dnub$, as visible \textbf{in} Figures~\ref{fig:cddiag} and~\ref{fig:ensemble}. We quantified this unexpected feature through the difference in $\langle \delta\nu_\mathrm{0 \ell} \rangle$ between RC and RGB stars, which is significantly different from one cluster to the other for both the small spacings. As discussed in Section~\ref{sec:unexp}, further theoretical investigations concerning differences on mass, metallicity and rotation between the two cluster populations, are required for the full interpretation of our results. 

\item The positions of the ridges in the collapsed ensemble \'{e}chelle diagrams (Figure~\ref{fig:ensemble}) confirm the results from \cite{Huber10redgiant}, with the $\ell = 1$ and $2$ ridges moving away from the $\ell = 0$ ridge as the stars evolve from the H-shell to the He-core burning phase. The position of the $\ell = 3$ ridges, hence of the average small spacings $\delta \nu_\mathrm{03}$ (Figures~\ref{fig:ensemble}(b) and (d)), is also in agreement with results by \cite{Bedding10redgiant,Huber10redgiant,Mosser11universal,K12} on field red giants. The FWHM of $\ell = 0$ ridge, which represents an upper limit of the mode linewidths, increases with $\Delta \nu$ for both NGC~6791 and NGC~6819, a result that agrees with the studies by \cite{Huber10redgiant} and by \cite{K12} on field red giants. A systematic difference of the FWHM between the two clusters is shown, which is largely explained by the temperature dependence of mode linewidths (Figure~\ref{fig:teff}), as discussed in Section~\ref{sec:fwhm}. This result shows the first evidence for an exponential correlation between mode linewidth and temperature in red giants, which is consistent with extrapolating \textit{Kepler} results for main-sequence and subgiant stars derived by \cite{App11temp}.

\item The analysis of period spacings using the method described by \cite{Bedding11gmode} allowed for the successful identification of almost half of the stars in our sample as either H-shell or He-core burning red giants. The fraction of RGB stars with clearly detectable period spacings is much lower than for RC stars, as seen in Figure~\ref{fig:dpobs}, and could be due to a weaker coupling of the p- and g-mode cavities in RGB stars. We see a number of outliers in Figure~\ref{fig:dpobs} which require further investigations. 
It is likely that all of them are binaries and two of them appear to be evolved blue stragglers, as suggested by \cite{RV98} and supported by their higher masses (Figure~\ref{fig:mass}). In addition, our analysis confirms the suggestion by \cite{Miglio11massloss} that stars KIC 2437103 ($\dP = 306 \, $s) and KIC 2437589 ($\dP = 39 \, $s), are an RC and an RGB star, respectively, and that KIC 5024404 ($\dP = 182 \, $s) is an RC star. Lastly, we find a number of possible candidates for evolved RC stars in both NGC~6791 and NGC~6819, as suggested by our measurement of their masses (Figure~\ref{fig:mass}), which indicate they have a radius larger than the other RC stars. The special cases discussed in Section~\ref{sec:special} represent potentially interesting targets for detailed theoretical modeling.
\end{enumerate}

\begin{deluxetable}{lccrccccrr}\tabletypesize{\footnotesize}
\rotate
\tablecolumns{9}
\tablewidth{0pc}
\tablecaption{Overall asteroseismic parameters for some interesting targets.} 

\tablehead{
\colhead{KIC ID} &
\colhead{NGC} &
\colhead{Notes\tablenotemark{a}} &
\colhead{$\nu_{\rm max}$ ($\mu$Hz)} &
\colhead{$\Delta \nu$ ($\mu$Hz)}  & 
\colhead{$\epsilon$} & 
\colhead{$\delta\nu_\mathrm{02}$ ($\mu$Hz)} &
\colhead{$\delta\nu_\mathrm{01}$ ($\mu$Hz)}  &   
\colhead{$\Delta P_\mathrm{obs}$ (s)}}
\startdata
5112361 & 6819 & (A) Outlier & $67.4 \pm 1.4$ & $6.181 \pm 0.025$ & $1.066 \pm 0.044$& $0.712 \pm 0.066$ & $-0.102 \pm 0.048$ & 75\\
4937770 & 6819 & (B) Outlier & $93.8 \pm 2.4$& $7.821 \pm 0.076$ & $1.119 \pm 0.117$ & $0.808 \pm 0.117$ & $-0.096 \pm 0.070$ & 109\\
5024414 & 6819 & (C) Outlier & $77.1 \pm 1.5$ & $6.490 \pm 0.056$ & $1.013 \pm 0.103$ & $0.720 \pm 0.072$ & $-0.220 \pm 0.143$ & 178\\
5024476 & 6819 & (D) Outlier & $67.0 \pm 1.7$ & $5.693 \pm 0.097$ & $1.138 \pm 0.201$ & $0.656 \pm 0.152$ & $-0.203 \pm 0.234$ & 199\\
2437103 & 6791 & Misclassified CMD & $29.7 \pm 1.7$ & $3.791 \pm 0.064$ & $0.770 \pm 0.132$ & $0.325 \pm 0.153$ & $-0.242 \pm 0.114$ & 306\\
2437589 & 6791 & Misclassified CMD & $46.5 \pm 1.5$ & $4.603 \pm 0.026$ & $1.026 \pm 0.057$ & $0.526 \pm 0.042$ & $-0.184 \pm 0.038$ & \multicolumn{1}{r}{39}\\
5024404 & 6819 & Misclassified CMD & $48.8 \pm 0.7$ & $4.857 \pm 0.126$ & $0.835 \pm 0.261$ & $0.689 \pm 0.095$ & $-0.122 \pm 0.072$ & 182\\
9716522 & 6811 & AGB & $54.9 \pm 1.0$ & $4.852 \pm 0.036$ & $0.973 \pm 0.084$ & $0.592 \pm 0.102$ & $-0.116 \pm 0.099$ & 154\\
2436417 & 6791 & Likely evolved RC & $26.7 \pm 0.8$ & $3.412 \pm 0.058$ & $0.874 \pm 0.133$ & $0.342 \pm 0.090$ & $-0.237 \pm 0.074$ & 268\\
2437804 & 6791 & Likely evolved RC & $26.5 \pm 1.6$ & $3.350 \pm 0.070$ & $0.870 \pm 0.165$ & $0.478 \pm 0.054$ & $-0.266 \pm 0.529$ & 212\\
5024601 & 6819 & Likely evolved RC & $31.8 \pm 1.7$ & $3.704 \pm 0.028$ & $0.862 \pm 0.065$ & $0.498 \pm 0.061$ & $-0.140 \pm 0.107$ & -\\
5112401 & 6819 & Likely evolved RC & $38.2 \pm 0.7$ & $4.047 \pm 0.068$ & $0.892 \pm 0.158$ & $0.476 \pm 0.082$ & $-0.169 \pm 0.069$ & 209\\
5112950 & 6819 & Likely evolved RC & $42.8 \pm 1.3$ & $4.302 \pm 0.036$ & $1.082 \pm 0.083$& $0.584 \pm 0.104$ & \multicolumn{1}{r}{$0.010 \pm 0.181$} & 249\\
5112974 & 6819 & Likely evolved RC & $41.7 \pm 0.7$ & $4.358 \pm 0.045$ & $0.874 \pm 0.099$ & $0.655 \pm 0.115$ & \multicolumn{1}{r}{$0.064 \pm 0.073$} & 239 &

\enddata
\tablenotetext{a}{Target description as presented in Section~\ref{sec:dpobs}.}
\label{tab:outliers}
\end{deluxetable} 

\acknowledgements{Funding for this Discovery mission is provided by NASAÕs Science Mission Directorate. The authors would like to thank the entire Kepler team, without whom this investigation would not have been possible. The research leading to these results has received funding from the European Community's Seventh Framework Programme (FP7/2007-2013) under grant agreement no. 269194. DS acknowledges support from the Australian Research Council. KB acknowledges funding from the Carlsberg Foundation. SH acknowledges financial support from the Netherlands Organisation for Scientific research (NWO). TK is supported by the FWO-Flanders under project O6260 - G.0728.11. NCAR is partially supported by the National Science Foundation. This research was supported in part by the National Science Foundation under Grant No. NSF PHY05-51164. Data presented in this paper are available upon request to the first author.}

\clearpage



\clearpage

\clearpage

\end{document}